\documentclass[final,3p,times]{elsarticle}

\usepackage{amssymb}
\usepackage{amsthm}

\usepackage{mathtools, float}
\usepackage{booktabs} 
\usepackage{multirow}
\usepackage{makecell}
\usepackage{multicol}
\usepackage{tabularx}
\usepackage{caption}
\usepackage{subcaption}
\usepackage{xcolor}

\usepackage[colorlinks=false, citecolor=red, linkcolor=blue, pdfborder={0 0 0}]{hyperref}

\usepackage[T1]{fontenc}
\usepackage{selinput}

\journal{Mechanical Systems and Signal Processing}

\begin{document}

\begin{frontmatter}



\title{On the use of Lagrange Multiplier State-Space Substructuring in Dynamic Substructuring analysis}


\author[inst1]{R.S.O. Dias \texorpdfstring{\corref{cor1}}{Lg}}
\cortext[cor1]{Corresponding author.}
\ead{r.dasilva@staff.univpm.it}
\affiliation[inst1]{organization={Department of Industrial Engineering and Mathematical Sciences, Universit\`a Politecnica delle Marche},
            addressline={Via Brecce Bianche}, 
            city={Ancona},
            postcode={60131}, 
            state={Marche},
            country={Italy}}

\author[inst1]{M. Martarelli}
\ead{m.martarelli@staff.univpm.it}
\author[inst2]{P. Chiariotti}
\ead{paolo.chiariotti@polimi.it}

\affiliation[inst2]{organization={Department of Mechanical Engineering, Politecnico di Milano},
            addressline={Via Privata Giuseppe La Masa}, 
            city={Milan},
            postcode={20156}, 
            state={Lombardia},
            country={Italy}}
            

\begin{abstract}
In this article, the formulation of Lagrange Multiplier State-Space Substructuring (LM-SSS) is presented and extended to directly compute coupled displacement and velocity state-space models. The LM-SSS method is applied to couple and decouple state-space models established in the modal domain. Moreover, it is used together with tailored post-processing procedures to eliminate the redundant states originated from the coupling and decoupling operations. This specific formulation of the LM-SSS approach made it possible to develop a tailored coupling form, named Unconstrained Coupling Form (UCF). UCF just requires the computation of a nullspace and does not rely on the selection of a subspace from a nullspace. An explanation of all the steps in order to compute state-space models without redundant states originated from the coupling and decoupling procedures is also given. \textcolor{black}{By exploiting a numerical example,} LM-SSS was compared with the Lagrange Multiplier Frequency Based Substructuring (LM-FBS) approach, which is currently widely recognized as a reference approach. This was done both in terms of: a) coupled FRFs derived by coupling the state-space models of two substructures and b) decoupled FRFs derived by decoupling the state-space model of a component from the coupled model. As for the first validation, LM-SSS showed to be suitable to compute minimal order coupled models and UCF turned out to have similar performance as other coupling forms already presented to the scientific community. As for the decoupling task, the FRFs derived from the LM-SSS approach turned out to perfectly match those obtained by LM-FBS. Moreover, it was also demonstrated that the elimination of the redundant states originated from the decoupling operation was correctly performed. \textcolor{black}{As final validation, the approaches discussed were exploited on an experimental substructuring application. LM-SSS resulted to be a reliable SSS technique to perform coupling and decoupling operations with state-space models estimated from measured FRFs as well as to provide accurate minimal-order models.}
\end{abstract}


\begin{highlights}
\item LM-SSS is reliable to couple/decouple SSMs estimated from measured FRFs. 
\item UCF performs similarly to the coupling forms available in literature.
\item UCF avoids the computation of multiple nullspaces and the selection of subspaces.
\item The coupled SSMs minimal form is found by using tailored post-processing procedures.
\item The unique set of DOFs can be retained by using a Boolean localization matrix.
\end{highlights}


\end{frontmatter}



\section{Introduction}\label{Introduction}

Dynamic Substructuring (DS) was introduced in the last century with the belief that complex structures can be analyzed in a more efficient way if they are considered as assemblies of several simpler components.  From its first introduction, this idea pushed the scientific community to an intense activity in this field, and several DS approaches have been released \textcolor{black}{\citep{DK_20081169}}. Examples of those techniques are the ones clustered under the labels Component Mode Synthesis (CMS) and Frequency Based Substructuring (FBS). In the CMS techniques (see, for instance \citep{RM_1968} and \citep{DJ_2004}), the substructures are usually represented in  modal domain by using a set of modes (which might include static, quasi-static and vibration modes depending on the CMS method \citep{DK_20081169}). These methods found many applications when a DS analysis is intended to be performed with numerical data.

In the FBS methods, the components are characterized by their Frequency Response Functions (FRFs). For this reason, FBS methods are typically used with experimental data. Examples of FBS techniques developed can be found, for instance, in \citep{BJ_198855} and \citep{DK06}. In \citep{DJ_2011}, a new DS technique denominated Impulse Based Substructuring (IBS) is discussed. This technique describes the substructures by using Impulse Response Functions (IRF). Hence, IBS uses a time domain formulation with respect to FBS for performing DS.

Another group of DS methods is the one named State-Space Substructuring (SSS). In SSS techniques the components are represented by state-space models. These methods are suitable to deal with both numerical and experimental data. The approach discussed in this paper belongs to this group of methods.

One of the first SSS techniques was developed by Su and Juang in \citep{SJ_94}. Their approach will be referred from now on as "classical SSS". The implementation of this method relies on the construction of an uncoupled state-space model by writing the state-space models of the substructures in a block diagonal form. Afterwards, a coupling matrix is defined and used to enforce the coupling conditions (compatibility and equilibrium) on this diagonal uncoupled model in order to obtain the coupled state-space model. This method is simple to understand, being also able to couple several structures at same time. Nevertheless, it has some important limitations. Firstly, it requires the inversion of two different matrices to compute the coupled state-space model, secondly it is unable to compute minimal-order coupled state-space models. This means that the computed coupled state-space model will contain redundant states. The presence of such states does not affect the quality of the input-output transfer functions that can be computed from the coupled state-space model \citep{SJ_94}. However, by computing a minimal-order state-space model the number of states is reduced and consequently the computational effort required in calculations with the coupled state-space model will be also reduced. On the one hand, this minimal-order representation paves the way to the use of these approaches for Hardware-in-the-loop applications (e.g. real-time substructuring), on the other hand, it turns out to be a more elegant representation of the problem. 

As we eliminate one DOF for each pair of connected DOFs, one state and its respective first order derivate must also be eliminated to avoid the presence of redundant states \citep{SJ_94}. Hence, the minimal-order state-space model will always present $n-2n_{J}$ states (being $n$ the sum of the number of states of the coupled state-space models and $n_{J}$ the number of interface outputs). If the coupling is performed with state-space models written in physical coordinates (for which the physical meaning of the states and outputs is the same), the states, which initially represented matching interface outputs, are representative of the correspondent coupled interface output (the same is valid for the respective first order derivatives). Hence, they represent the same physical quantity. However, the computation of the minimal-order coupled model cannot be performed by simply keeping on the coupled model just one state from each group of states (whose physical meaning is the same), because each of these states has a non-negligible contribution for the state-space model. Therefore, this state elimination must guarantee that the contribution of the states to be eliminated is retrieved by the one that is kept on the model. To properly perform the elimination of the redundant states, one can follow the post-processing procedures described in \citep{RD_2021}.

When working with state-space models estimated from measured data, these models are usually obtained in modal domain. Hence, it is difficult to say which states correspond to matching interface outputs, turning the elimination of the redundant states infeasible. On the other hand, as pointed out by Gibanica et al. in \citep{mg_2013},\citep{MG_2014}, a transformation of a state-space model written in modal coordinates into physical ones is not suitable for experimental models. In general, for these models $n >> n_{o}$ (being $n_{o}$ the number of outputs), hence this transformation would impose a tremendous constraint on the model in order to force it to have as many states as two times the number of outputs. 

A solution to obtain minimal-order coupled state-space models was outlined by Sj\"ovall and Abrahamsson in \citep{SJO_20072697}. They suggested to partially transform the state vector of the state-space models of the substructures into physical domain before coupling them. This transformation was denominated as coupling form being enforced by applying a similarity state vector transformation as described in \citep{SJO_20072697}. By performing this procedure, the state vector of the state-space models will be composed by the first derivate of the interface outputs, by the interface outputs and $n-2n_{J}$ internal states.

When the state-space models to be coupled are transformed into this coupling form, their state-space matrices present a particular structure, for which the first block row of the state equation is representative of a second order differential equation, making it possible to perform coupling by directly summing those block rows \citep{mg_2013},\citep{MEL_2019325}. This coupling procedure is possible, because the applied similarity transformation guarantees that the contribution of the interface inputs to the response of the internal states is null. Therefore, the internal states remain uncoupled and are just correctly placed on the coupled state-space model \citep{mg_2013}. This SSS technique is able to mitigate the problems of classical SSS, since the coupling is made by performing just one matrix inversion and the coupled model is directly obtained in a minimal-order form. However, it introduces the disadvantage of coupling a maximum of two substructures at same time. Furthermore, to compute the transformation matrix to transform a state-space model into coupling form the choice of a subspace from a nullspace is required. This choice is hard to make and strongly affects the performance of the transformation \citep{mg_2013}. Recently, a new coupling form that requires the computation of two nullspaces, but does not rely on the choice of any subspace was proposed in \textcolor{black}{\citep[section~4.9.4]{AMRDvMTPATMR_2020}} and its performance will be also evaluated in this document.  

Gibanica in \citep{mg_2013} ported the technique developed by  Sj\"ovall and Abrahamsson in \citep{SJO_20072697} into the general framework presented by de Klerk et al. \citep{DK_20081169}. By performing such porting operation, the limitation of coupling a maximum of two substructures at same time was eliminated, but two matrix inversions were now required to compute the coupled state-space model.

\textcolor{black}{Recently, a novel SSS method was introduced by Kammermeier et al. in \citep{BK_2020}. The same method was later named as Lagrange Multiplier State-Space Substructuring (LM-SSS) method by Dias et al. in \citep{RD_2021}. LM-SSS just requires a single matrix inversion to compute the coupled state-space model, being able to couple several structures at same time. However, to compute minimal-order coupled models, post-processing procedures need to be applied as discussed in \citep{RD_2021}. Furthermore, as LM-SSS is a dual assembly formulation, the full set of interface DOFs will be retained during coupling or decoupling operations \citep{DK_20081169}. Thus, the elimination of the redundant DOFs from the state-space models computed by LM-SSS is also required.}

\textcolor{black}{With respect to the previous paper issued by the authors on LM-SSS \citep{RD_2021}, this paper aims at introducing the following novel elements to the method:
\begin{itemize}
    \item provide a solution to retain the unique set of DOFs from the coupled state-space models without relying on a manual elimination;
    \item extend LM-SSS to directly compute displacement and velocity state-space models;
    \item present a new coupling form, specifically tailored to LM-SSS, to ease the computation of minimal-order models when dealing with state-space models estimated from measured data;
    \item validate the whole approach numerically and experimentally.
\end{itemize}}

The LM-SSS method is described in section \ref{LM_SSS}, while the procedure to perform decoupling is presented in section \ref{decoupling}. Section \ref{Retaining the unique set of interface DOFs} presents a solution to retain the unique set of DOFs from the coupled state-space model. Then, in sections \ref{Unconstrained_Coupling_Form} and \ref{Minimal-order coupled state-space models} a new coupling form is introduced and the procedure to eliminate redundant states which are originated from coupling and decoupling operations is discussed, respectively. \textcolor{black}{In section \ref{Comparison of LM-SSS with other SSS techniques} a comparison between LM-SSS and the SSS methods developed by Su and Juang \citep{SJ_94} and Sj\"oval and Abrahamsson \citep{SJO_20072697} is given.} Afterwards, in section \ref{Numerical example} the performance of LM-SSS to couple and decouple state-space models established in modal domain is evaluated and discussed by using a numerical example, \textcolor{black}{while in section \ref{Experimental validation} the same analysis is performed on experimental data.} Finally, the conclusions are presented in section \ref{Conclusion}.

\section{Nomenclature}

For easier understanding of all the parameters used in this article, the nomenclature is given in table \ref{table:Nomenclature}.

\begin{table}[ht]
\centering
\caption{Nomenclature}
\begin{tabular}{@{}llll@{}}
\toprule
$A$ & state matrix & $M$ & mass matrix \\
$B$ & input matrix & $N$ & nullspace of a matrix\\
$B_{M}$ & mapping matrix & $q$ & state vector written in modal coordinates\\
\textcolor{black}{$B_{T}$} & \textcolor{black}{state mapping matrix} & $T$ & transformation matrix \\
$C$ & output matrix & $T_{SJ}$ & coupling matrix\\
$D$ & feed-through matrix & $u$ & input vector\\
$g$ & connecting forces vector & $V$ & damping matrix\\
$H$ & FRF matrix & $x$ & state vector\\
$K$ & stiffness matrix & \textcolor{black}{$x_{min}$} & \textcolor{black}{state vector of a minimal-order coupled model}\\
$L$ & Boolean localization matrix & $y$ & output vector\\
\textcolor{black}{$L_{T}$} & \textcolor{black}{state Boolean localization matrix} & $z$ & state vector transformed into coupling form\\
\vspace{1mm} & & &\\
$\gamma$ & Gaussian distributed independent stochastic variable & $\lambda$ & Lagrange multipliers vector\\ 
$\theta$ & Gaussian distributed independent stochastic variable &  $\mu$ & residual difference\\ 
\vspace{1mm} & & &\\
$\bullet_{accel}$             & acceleration state-space model & $\bullet_{j}$              & j:th input\\
$\bullet_{D}$            & block diagonal matrix & $\bullet_{k}$              & k:th discrete frequency \\
$\bullet_{disp}$               & displacement state-space model & $\bullet_{S}$             & S:th substructure\\
$\bullet_{i}$              & i:th output & $\bullet_{vel}$       & velocity state-space model\\
$\bullet_{\alpha}$ & substructure $\alpha$ & $\bullet_{\beta}$ & substructure $\beta$ \\
\vspace{1mm} & & &\\
$\bullet^{-1}$            & inverse of a matrix & $\bullet^{\dag}$  & pseudoinverse of a matrix\\ 
$\bullet^{I}$ &   internal DOF & $\bullet^{T}$              & transpose of a vector/matrix\\
$\bullet^{J}$ & interface DOF & &\\
\vspace{1mm} & & &\\
$\dot{\bullet}$  & first order time derivative & $\ddot{\bullet}$  & second order time derivative\\
$\bar{\bullet}$  & coupling vector/matrix & $\widetilde{\bullet}$  & vector composed by the unique set of DOFs\\ \midrule
\label{table:Nomenclature}
\end{tabular}
\end{table}

\section{Lagrange Multiplier State-Space Substructuring}\label{LM_SSS}

\begin{figure}[ht]
\centering
    \includegraphics[width=0.45\textwidth]{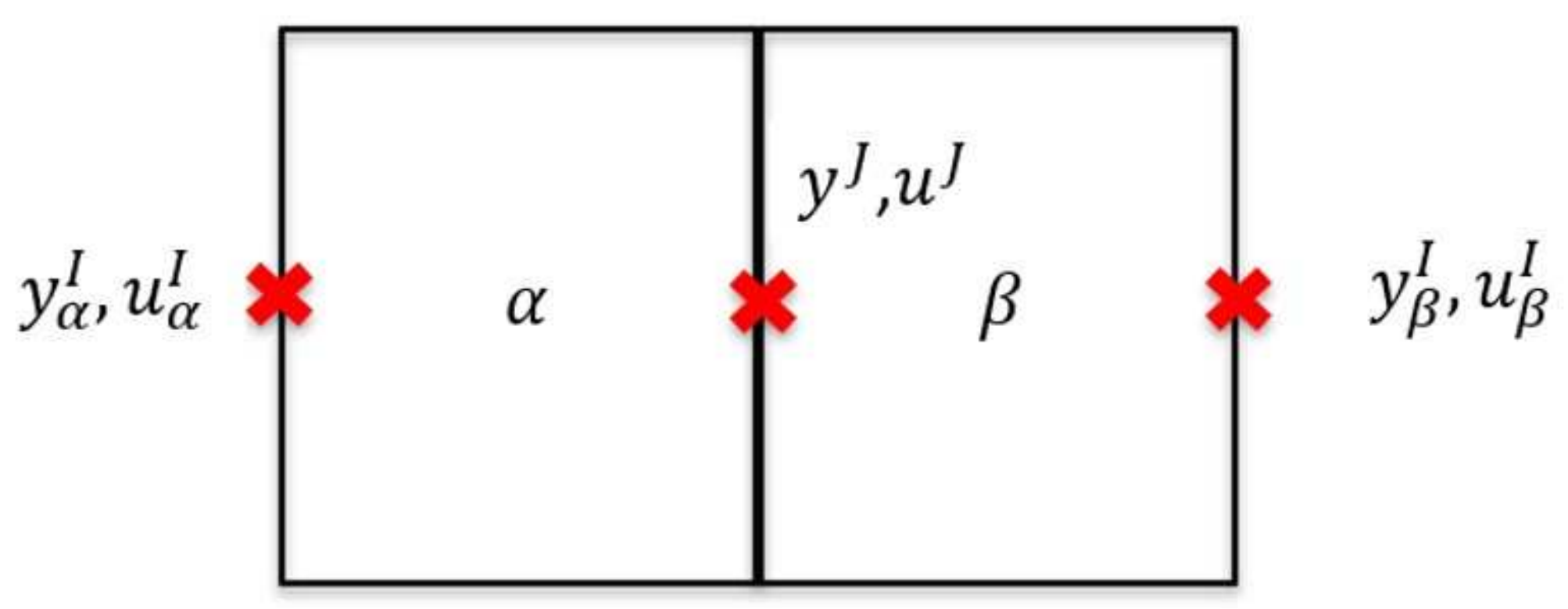}
    \caption{Assembled structure composed by two different components \citep{RD_2021}.}
     \label{fig:Assembly_rigid_coupling}
\end{figure}

The Lagrange Multiplier State-Space Substructuring is a novel SSS method introduced in \citep{BK_2020}. To get a short insight into the approach, let us consider the assembled structure shown in figure \ref{fig:Assembly_rigid_coupling}.

\begin{figure}[ht]
\centering
    \includegraphics[width=0.6\textwidth]{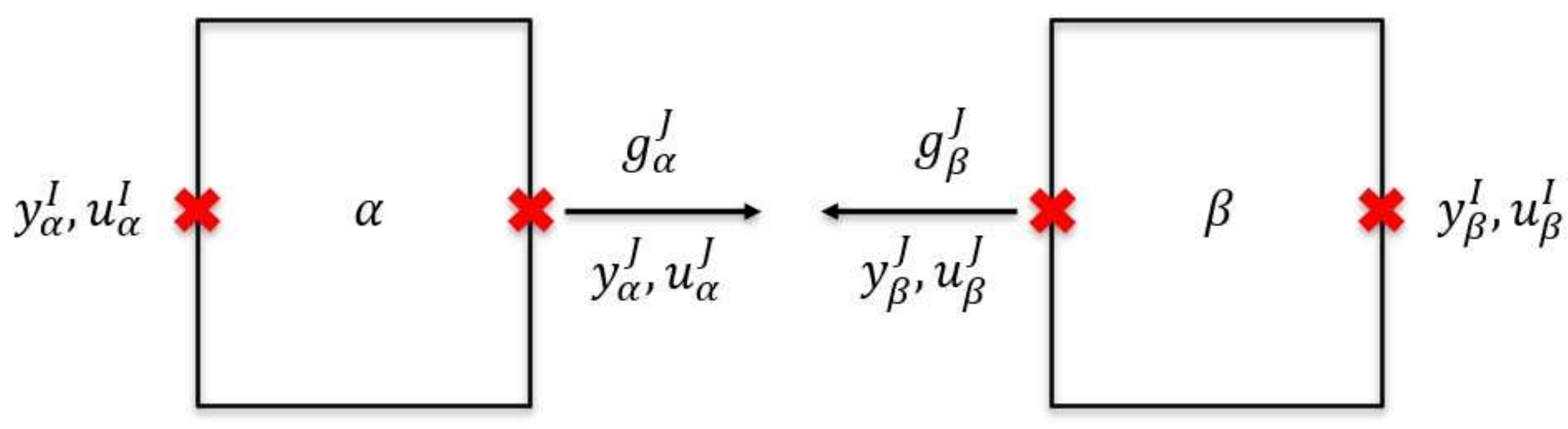}
    \caption{Components virtually separated \citep{RD_2021}.}
     \label{fig:Free_Body_Diagram_Rigid_Coupling}
\end{figure}

LM-SSS relies on the construction of a diagonal coupled state-space model. This model is obtained by writing the state-space matrices of each substructure to be coupled in block diagonal form and by using the forces that are responsible to keep the substructures coupled, which are usually tagged as connecting forces. When the substructures are coupled (see figure \ref{fig:Assembly_rigid_coupling}) the connecting forces that are acting at their interfaces are mutually cancelled, since they are equal in intensity and opposite in direction. However, if these substructures are virtually separated, we may observe the connecting forces acting at the interface of each component as depicted in figure \ref{fig:Free_Body_Diagram_Rigid_Coupling}.  

As $\{g^{J}_{\alpha}(t)\}=-\{g^{J}_{\beta}(t)\}$, we may use the mapping matrix presented by de Klerk et al. in \citep{DK06} in order to obtain a more compact representation for the relation between both connecting forces:

\begin{equation}\label{eq:g_lambda}
\left\{\begin{matrix}
\{g_{\alpha}(t)\}\\
\{g_{\beta}(t)\}
\end{matrix}\right\}=-[\begin{matrix}
B_{M}
\end{matrix}]^{T}\{\lambda(t)\}
\end{equation}

where, $\left\{ \lambda(t) \right\}$ is the vector of Lagrange Multipliers that represent connecting forces \citep{DK06} and $[B_{M}]$ represents the signed Boolean mapping matrix. Note that, this matrix is presented as $[B]$ matrix in \citep{DK06}, but here will be denoted as $[B_{M}]$ to avoid confusion with the $[B]$ state-space matrix. The vectors $\{g_{\alpha}(t)\}$ and $\{g_{\beta}(t)\}$ are given as follows.

    \begin{subequations}\label{zero4}
        \noindent
        \begin{tabularx}{\linewidth}{XX}
        \begin{equation}
        \{g_{\alpha}(t)\}=\left\{\begin{matrix}
        \{0\}\\
        \{g^{J}_{\alpha}(t)\}
        \end{matrix}
        \right\} \label{three}
        \end{equation}
        &
        \begin{equation}
        \{g_{\beta}(t)\}=\left\{\begin{matrix}
        \{0\}\\
        \{g^{J}_{\beta}(t)\}
        \end{matrix}
        \right\} \label{four}
        \end{equation}
    \end{tabularx}
    \end{subequations}

Considering figure \ref{fig:Free_Body_Diagram_Rigid_Coupling} and equation \eqref{eq:g_lambda} and generalizing for coupling an unlimited number of structures, a diagonal coupled state-space model can be constructed as follows

\begin{equation}\label{eq:simplified_coupled_ss_model_from_uncoupled_ss_0}
\begin{gathered}
\left\{\begin{matrix}
\{\dot{x}_{\alpha}(t)\}\\
\{\dot{x}_{\beta}(t)\}\\
\vdots
\end{matrix}
\right\}=[A_{D}]\left\{\begin{matrix}
\{x_{\alpha}(t)\}\\
\{x_{\beta}(t)\}\\
\vdots
\end{matrix}
\right\}+[B_{D}]\left(\left\{\begin{matrix}
\{u_{\alpha}(t)\}\\
\{u_{\beta}(t)\}\\
\vdots
\end{matrix}
\right\}-[B_{M}]^{T}\left\{\begin{matrix}
\{\lambda_{\alpha}(t)\}\\
\{\lambda_{\beta}(t)\}\\
\vdots
\end{matrix}
\right\}\right)\\
\left\{\begin{matrix}
\{\ddot{y}_{\alpha}(t)\}\\
\{\ddot{y}_{\beta}(t)\}\\
\vdots
\end{matrix}
\right\}=[C_{accel,D}]\left\{\begin{matrix}
\{x_{\alpha}(t)\}\\
\{x_{\beta}(t)\}\\
\vdots
\end{matrix}
\right\}+[D_{accel,D}]\left(\left\{\begin{matrix}
\{u_{\alpha}(t)\}\\
\{u_{\beta}(t)\}\\
\vdots
\end{matrix}
\right\}-[B_{M}]^{T}\left\{\begin{matrix}
\{\lambda_{\alpha}(t)\}\\
\{\lambda_{\beta}(t)\}\\
\vdots
\end{matrix}
\right\}\right)
\end{gathered}
\end{equation}

where,

\begin{alignat}{2}\label{eq:MatrixABCD}
\begin{split}
[A_{D}]=\left[
\begin{matrix}
A_{\alpha} & &\\
 & A_{\beta} &\\
 & & \ddots
\end{matrix}
\right],\ \ \ & [B_{D}]=\left[
\begin{matrix}
B_{\alpha} & &\\
 & B_{\beta} &\\
 & & \ddots
\end{matrix}
\right]\\ 
[C_{accel,D}]=\left[\begin{matrix}
C_{accel,\alpha} & &\\
 & C_{accel,\beta} &\\
 & & \ddots
\end{matrix}
\right],\ \ \ & [D_{accel,D}]=\left[
\begin{matrix}
D_{accel,\alpha} & &\\
 & D_{accel,\beta} &\\
 & & \ddots
\end{matrix}
\right]
\end{split}
\end{alignat}

$\left\{ x(t) \right\} \in \mathbb{R}^{n \times 1}$ represents the state vector, $\left\{ u(t) \right\} \in \mathbb{R}^{n_{i} \times 1}$ represents the input vector, whose elements are forces and $\left\{\ddot{y}(t) \right\} \in \mathbb{R}^{n_{o} \times 1}$ represents the acceleration output vector. Subscripts $\alpha$ and $\beta$ denote variables related to substructures $\alpha$ and $\beta$, respectively. Subscript $accel$ denotes a state-space matrix of an acceleration state-space model (state-space model whose output vector elements are accelerations), while subscript $D$ denotes a block diagonal matrix.

By using a more compact representation, the diagonal coupled state-space model (equation \eqref{eq:simplified_coupled_ss_model_from_uncoupled_ss_0}) can be rewritten as follows:

\begin{equation}\label{eq:simplified_coupled_ss_model_from_uncoupled_ss}
\begin{gathered}
\{\dot{x}(t)\}=[A_{D}]\{x(t)\}+[B_{D}](\{u(t)\}-[B_{M}]^{T}\{\lambda(t)\})\\
\{\ddot{y}(t)\}=[C_{accel,D}]\{x(t)\}+[D_{accel,D}](\{u(t)\}-[B_{M}]^{T}\{\lambda(t)\})
\end{gathered}
\end{equation}

When substructures are coupled, they must verify continuity at their interface. This condition can be mathematically represented by using the mapping matrix $[B_{M}]$ as follows:

\begin{equation}\label{eq:compatibility_condition0}
[B_{M}]\{y(t)\}=\{0\}
\end{equation}

Since we are working with state-space models whose outputs represent accelerations (acceleration state-space models), equation \eqref{eq:compatibility_condition0} must hold also for accelerations.

\begin{equation}\label{eq:compatibility_condition}
[B_{M}]\{\ddot{y}(t)\}=\{0\}
\end{equation}

Besides the interface continuity requirement, coupled substructures must also verify the local equilibrium equations given by the output equations of the diagonal coupled state-space model (equation \eqref{eq:simplified_coupled_ss_model_from_uncoupled_ss}). By solving these equations it is possible to find the value of $\{u(t)\}$:

\begin{equation}\label{eq:equilibrium_condition_rearranged}
\{u(t)\}=[D_{accel,D}]^{-1}(\{\ddot{y}(t)\}-[C_{accel,D}]\{x(t)\})+[B_{M}]^{T}\{\lambda(t)\}
\end{equation}

By using together equations \eqref{eq:compatibility_condition} and \eqref{eq:equilibrium_condition_rearranged}, and dropping $\{\bullet\}$, $[\bullet]$ and $(t)$ for ease of readability, we can write these equations as given below:

\begin{equation}\label{eq:coupling_conditions}
\begin{cases}
(D_{accel,D})^{-1}(\ddot{y}-C_{accel,D}x)+B_{M}^{T}\lambda=u\\
B_{M}\ddot{y}=0
\end{cases}
\end{equation}

From the system of equations \eqref{eq:coupling_conditions}, the following relations can be obtained:

\begin{equation}\label{eq:coupling_conditions_1}
\begin{cases}
\lambda=(B_{M}D_{accel,D}B_{M}^{T})^{-1}(B_{M}C_{accel,D}x+B_{M}D_{accel,D}u)\\
\ddot{y}=(C_{accel,D}-D_{accel,D}B_{M}^{T}(B_{M}D_{accel,D}B_{M}^{T})^{-1}B_{M}C_{accel,D})x+(D_{accel,D}-D_{accel,D}B_{M}^{T}(B_{M}D_{accel,D}B_{M}^{T})^{-1}B_{M}D_{accel,D})u
\end{cases}
\end{equation}

By using equation \eqref{eq:equilibrium_condition_rearranged} and the bottom equation of the system of equations \eqref{eq:coupling_conditions_1}, the state equations of the diagonal coupled state-space model (equation \eqref{eq:simplified_coupled_ss_model_from_uncoupled_ss}) can be rewritten as follows:

\begin{equation}\label{eq:coupled_state_equation_final}
\dot{x}=(A_{D}-B_{D}B_{M}^{T}(B_{M}D_{accel,D}B_{M}^{T})^{-1}B_{M}C_{accel,D})x+(B_{D}-B_{D}B_{M}^{T}(B_{M}D_{accel,D}B_{M}^{T})^{-1}B_{M}D_{accel,D})u
\end{equation}

By using equation \eqref{eq:coupled_state_equation_final} and the bottom equation of the system of equations \eqref{eq:coupling_conditions_1}, the coupled state-space model is achieved as follows:

\begin{equation}\label{eq:diagonalfirstorderformAB}
\begin{gathered}
\{\dot{\bar{x}}(t)\}=[\bar{A}]\{x(t)\}+[\bar{B}]\{\bar{u}(t)\}\\
\{\ddot{\bar{y}}(t)\}=[\bar{C}_{accel}]\{\bar{x}(t)\}+[\bar{D}_{accel}]\{\bar{u}(t)\}
\end{gathered}
\end{equation}

where,

\begin{equation}\label{eq:coupled_matrices_value}
\begin{gathered}
[\bar{A}]=A_{D}-B_{D}B_{M}^{T}(B_{M}D_{accel,D}B_{M}^{T})^{-1}B_{M}C_{accel,D}\\
[\bar{B}]=B_{D}-B_{D}B_{M}^{T}(B_{M}D_{accel,D}B_{M}^{T})^{-1}B_{M}D_{accel,D}\\
[\bar{C}_{accel}]=C_{accel,D}-D_{accel,D}B_{M}^{T}(B_{M}D_{accel,D}B_{M}^{T})^{-1}B_{M}C_{accel,D}\\
[\bar{D}_{accel}]=D_{accel,D}-D_{accel,D}B_{M}^{T}(B_{M}D_{accel,D}B_{M}^{T})^{-1}B_{M}D_{accel,D}
\end{gathered}
\end{equation}

overbar variables represent variables of the coupled state-space model.

At this point, is important to mention that several system identification methods developed to estimate state-space models from FRFs provide an estimation of the displacement state-space model (see for example \citep{MG_2020}). However, to deduce LM-SSS technique we have used acceleration state-space models, hence this SSS technique has been established just to couple this kind of models. Thus, in general the estimated state-space models must be double differentiated in order to obtain the respective acceleration model.

By double differentiating a generic displacement state-space model, the respective acceleration state-space model is obtained as follows \citep{FL_1988}:

\begin{equation}\label{eq:acceleration_ss_model}
\begin{gathered}
\{\dot{x}(t)\}=[A]\{x(t)\}+[B]\{u(t)\}\\
\{\ddot{y}(t)\}=[C_{disp}AA]\{x(t)\}+([D_{disp}]+[C_{disp}AB])\{u(t)\}
\end{gathered}
\end{equation}

Observing equation \eqref{eq:acceleration_ss_model}, we conclude that the acceleration state-space matrices can be calculated from the displacement ones as follows:

\begin{equation}\label{eq:acceleration_ss_matrices}
[A_{accel}]=\left[\begin{matrix}
A
\end{matrix}
\right],\ \ \
 [B_{accel}]=\left[\begin{matrix}
B
\end{matrix}
\right],\ \ \
[C_{accel}]=\left[\begin{matrix}
C_{disp}AA
\end{matrix}
\right]\ \ \
[D_{accel}]=\left[\begin{matrix}
C_{disp}AB
\end{matrix}
\right]+[D_{disp}]
\end{equation}

By using the expressions in equation \eqref{eq:acceleration_ss_matrices}, we may rewrite equations \eqref{eq:coupled_matrices_value} as follows:

\begin{equation}\label{eq:coupled_matrices_value_from_disp}
\begin{gathered}
[\bar{A}]=A_{D}-B_{D}B_{M}^{T}(B_{M}(C_{disp,D}A_{D}B_{D})B_{M}^{T})^{-1}B_{M}(C_{disp,D}A_{D}A_{D})\\
[\bar{B}]=B_{D}-B_{D}B_{M}^{T}(B_{M}(C_{disp,D}A_{D}B_{D})B_{M}^{T})^{-1}B_{M}(C_{disp,D}A_{D}B_{D})\\
[\bar{C}_{accel}]=C_{disp,D}A_{D}A_{D}-(C_{disp,D}A_{D}B_{D})B_{M}^{T}(B_{M}(C_{disp,D}A_{D}B_{D})B_{M}^{T})^{-1}B_{M}(C_{disp,D}A_{D}A_{D})\\
[\bar{D}_{accel}]=C_{disp,D}A_{D}B_{D}-(C_{disp,D}A_{D}B_{D})B_{M}^{T}(B_{M}(C_{disp,D}A_{D}B_{D})B_{M}^{T})^{-1}B_{M}(C_{disp,D}A_{D}B_{D})
\end{gathered}
\end{equation}

By using expressions \eqref{eq:acceleration_ss_matrices}, equation \eqref{eq:coupled_matrices_value_from_disp} can be rewritten in order to directly provide the coupled displacement state-space model:

\begin{equation}\label{eq:disp_coupled_matrices_value_from_disp}
\begin{gathered}
[\bar{A}]=A_{D}-B_{D}B_{M}^{T}(B_{M}(C_{disp,D}A_{D}B_{D})B_{M}^{T})^{-1}B_{M}(C_{disp,D}A_{D}A_{D})\\
[\bar{B}]=B_{D}-B_{D}B_{M}^{T}(B_{M}(C_{disp,D}A_{D}B_{D})B_{M}^{T})^{-1}B_{M}(C_{disp,D}A_{D}B_{D})\\
[\bar{C}_{disp}]=C_{disp,D}-(C_{disp,D}A_{D}B_{D})B_{M}^{T}(B_{M}(C_{disp,D}A_{D}B_{D})B_{M}^{T})^{-1}B_{M}(C_{disp,D})\\
[\bar{D}_{disp}]=C_{disp,D}A_{D}B_{D}-(C_{disp,D}A_{D}B_{D})B_{M}^{T}(B_{M}(C_{disp,D}A_{D}B_{D})B_{M}^{T})^{-1}B_{M}(C_{disp,D}A_{D}B_{D})-\bar{C}_{disp}\bar{A}\bar{B}=0
\end{gathered}
\end{equation}

\textcolor{black}{From the mathematical relations between an acceleration and a velocity state-space model (see \citep{FL_1988}), expression \eqref{eq:coupled_matrices_value_from_disp} can be rewritten to directly compute the coupled velocity state-space model as follows:}

\begin{equation}\label{eq:vel_coupled_matrices_value_from_disp}
\begin{gathered}
[\bar{A}]=A_{D}-B_{D}B_{M}^{T}(B_{M}(C_{disp,D}A_{D}B_{D})B_{M}^{T})^{-1}B_{M}(C_{disp,D}A_{D}A_{D})\\
[\bar{B}]=B_{D}-B_{D}B_{M}^{T}(B_{M}(C_{disp,D}A_{D}B_{D})B_{M}^{T})^{-1}B_{M}(C_{disp,D}A_{D}B_{D})\\
[\bar{C}_{vel}]=C_{disp,D}A_{D}-(C_{disp,D}A_{D}B_{D})B_{M}^{T}(B_{M}(C_{disp,D}A_{D}B_{D})B_{M}^{T})^{-1}B_{M}(C_{disp,D}A_{D})\\
[\bar{D}_{vel}]=C_{disp,D}B_{D}-(C_{disp,D}A_{D}B_{D})B_{M}^{T}(B_{M}(C_{disp,D}A_{D}B_{D})B_{M}^{T})^{-1}B_{M}(C_{disp,D}B_{D})
\end{gathered}
\end{equation}

where, subscrip $vel$ denotes a state-space matrix of a velocity state-space model. If the coupled state-space model respects the second law of Newton $[C_{disp,D}B_{D}]=[0]$ \citep{FL_1988},\citep{SJO_20072697}, hence $[\bar{D}_{vel}]=[0]$.

\section{Decoupling}\label{decoupling}

Decoupling is used to identify the dynamic behaviour of a component that is part of an assembled structure. This approach can only be applied if the dynamic behaviour of the assembled structure and of the remaining components is known. By using decoupling we may benefit from several advantages, for instance, having the possibility to dynamically characterize parts that are difficult to be tested separately \citep{WA_2012} (such as rubber mounts \citep{MH_2020},\citep{JO_2020}). 

To decouple a substructure we may use the Inverse Coupling approach \citep{MS_2015},\citep{WA_2012}. This decoupling procedure follows the same methodology as coupling. However, the state-space model of the substructure to be decoupled must be established in negative form \citep{MEL_2019325},\citep{MS_2015}. Then from the obtained state-space model, the inputs and outputs of the substructure to be characterized must be retained while the others must be eliminated. This is equivalent to just keep the columns of the input and feed-through matrices that are being multiplied by the inputs of the substructure to be identified and keep the rows of the output and feed-through matrices correspondent to the outputs of the same substructure.

The negative form of a state-space model can be derived by using a numerical state-space model. This kind of models can be directly computed from the mass, stiffness and damping matrices of the mechanical system being analyzed \citep{mg_2013}. Let us consider the following acceleration state-space model:

\begin{equation}\label{eq:accel_ss_model}
\begin{gathered}
\{\dot{x}(t)\}=[A]\{x(t)\}+[B]\{u(t)\}\\
\{\ddot{y}(t)\}=[C]\{x(t)\}+[D]\{u(t)\}
\end{gathered}
\end{equation}

where, the state-space matrices are given as follows:

\begin{alignat}{2}\label{eq:MatrixABCD_numerical}
\begin{split}
[A]=\left[
\begin{matrix}
-M^{-1}V & -M^{-1}K\\
I & 0\\
\end{matrix}
\right],\ \ \ & [B]=\left[
\begin{matrix}
M^{-1}\\
0
\end{matrix}
\right]\\ 
[C]=\left[\begin{matrix}
-M^{-1}V & -M^{-1}K
\end{matrix}
\right],\ \ \ & [D]=\left[
\begin{matrix}
M^{-1}
\end{matrix}
\right]
\end{split}
\end{alignat}

matrices $[M]$, $[K]$ and $[V]$ are the mass, stiffness and damping matrices respectively. 

The computation of the negative form of the state-space model given by expression \eqref{eq:accel_ss_model} can be performed by multiplying the matrices $[M]$, $[K]$ and $[V]$ by $-1$ \citep{MEL_2019325},\citep{MS_2015}, as follows:

\begin{align}\label{eq:negativestatespacemodel}
\begin{split}
[A]=\left[
\begin{matrix}
-(-M)^{-1}(-V) & -(-M)^{-1}(-K)\\
I & 0
\end{matrix}
\right],\ \ \  & [B]=\left[
\begin{matrix}
-M^{-1}\\
0
\end{matrix}
\right]\\
[C]=[
\begin{matrix}
-(-M)^{-1}(-V) & -(-M)^{-1}(-K)
\end{matrix}
],\ \ \ &  [D]=\left[
\begin{matrix}
-M^{-1}
\end{matrix}
\right]
\end{split}
\end{align}

By observing the state-space matrices given by expression \eqref{eq:negativestatespacemodel}, one realizes that to obtain the negative form of an acceleration state-space model, its input and feed-through matrices must be multiplied by $-1$, whereas the remaining ones do not go through any mathematical operation. Even though, the computation of the negative form of a state-space model was demonstrated for a theoretical model, this procedure continues to be valid when working with state-space models established in modal domain. 

\section{Retaining the unique set of interface DOFs}\label{Retaining the unique set of interface DOFs}

\textcolor{black}{The LM-SSS method is a dual DS formulation. These kind of DS formulations are characterized by retaining the full set of interface DOFs. This DOFs retention happens, because both established compatibility and equilibrium conditions (see equations \eqref{eq:compatibility_condition} and \eqref{eq:equilibrium_condition_rearranged}, respectively) do not provide a direct relation between the unique set of coupled interface outputs and inputs and the set of outputs and inputs of the state-space models to be coupled.} Hence, for each pair of connected DOFs, two outputs and inputs with the same physical meaning will be present on the coupled state-space model. The elimination of these repeated inputs and outputs can be easily achieved by eliminating one of the correspondent columns of the $[B]$ and $[D]$ matrices and one of the correspondent rows of the $[C]$ and $[D]$ matrices, respectively \citep{RD_2021}. However, when several DOFs are being coupled this procedure might get cumbersome, which may lead to erroneous eliminations. 

To easily eliminate the redundant inputs and outputs, the relations between the full sets and unique sets of inputs and outputs will be used. Those relations were established in \citep{DK_20081169},\textcolor{black}{\citep{MVS_2016}}, by using a Boolean localization matrix $[L]$, which can be computed from the nullspace of the mapping matrix $[B_{M}]$. These relations are given as follows:

    \begin{subequations}\label{zero3}
        \noindent
        \begin{tabularx}{\linewidth}{XX}
        \begin{equation}
        \{\widetilde{u}(t)\}=[L]^{T}\{u(t)\} \label{eq:L_full_unique_forces}
        \end{equation}
        &
        \begin{equation}
        \{y(t)\}=[L]\{\widetilde{y}(t)\} \label{eq:L_full_unique_displ}
        \end{equation}
    \end{tabularx}
    \end{subequations}

where, $\{\widetilde{u}(t)\}$ represents the unique set of force inputs and $\{\widetilde{y}(t)\}$ represents the unique set of displacements outputs.

Consider now a generic acceleration coupled state-space model obtained by using the LM-SSS method:

\begin{equation}\label{eq:generic_coupled_ss}
\begin{gathered}
\{\dot{\bar{x}}(t)\}=[\bar{A}]\{\bar{x}(t)\}+[\bar{B}]\{\bar{u}(t)\}\\
\{\ddot{\bar{y}}(t)\}=[\bar{C}]\{\bar{x}(t)\}+[\bar{D}]\{\bar{u}(t)\}
\end{gathered}
\end{equation}

Since the state-space model used here to demonstrate the retention of the unique set of DOFs by using the $[L]$ matrix presents accelerations as outputs, the second order time derivative of equation \eqref{eq:L_full_unique_displ} must be computed as follows:

\begin{equation}\label{eq:L_full_unique_accel}
\{\ddot{\bar{y}}(t)\}=[L]\{\ddot{\widetilde{y}}(t)\}
\end{equation}

By using equations \eqref{eq:L_full_unique_forces} and \eqref{eq:L_full_unique_accel}, expression \eqref{eq:generic_coupled_ss} can be rewritten as follows:

\begin{equation}\label{eq:retaining_DOFs_1}
\begin{gathered}
\{\dot{\bar{x}}(t)\}=[\bar{A}]\{\bar{x}(t)\}+[\bar{B}]([L]^{T})^{\dag}\{\widetilde{u}(t)\}\\
[L]\{\ddot{\widetilde{y}}(t)\}=[\bar{C}]\{\bar{x}(t)\}+[\bar{D}]([L]^{T})^{\dag}\{\widetilde{u}(t)\}
\end{gathered}
\end{equation}

where, superscript $\dag$ represents the pseudoinverse of a matrix.

By using the Moore-Penrose pseudoinverse of matrix $[L]$, the output equation of the state-space model given by expression \eqref{eq:retaining_DOFs_1} can be rewritten as follows:

\begin{equation}\label{eq:retaining_DOFs_2}
\begin{gathered}
\{\dot{\bar{x}}(t)\}=[\bar{A}]\{\bar{x}(t)\}+[\bar{B}]([L]^{T})^{\dag}\{\widetilde{u}(t)\}\\
\{\ddot{\widetilde{y}}(t)\}=[L]^{\dag}[\bar{C}]\{\bar{x}(t)\}+[L]^{\dag}[\bar{D}]([L]^{T})^{\dag}\{\widetilde{u}(t)\}
\end{gathered}
\end{equation}

Equation \eqref{eq:retaining_DOFs_2} represents the state-space model given by expression \eqref{eq:generic_coupled_ss} composed by a unique set of interface inputs and outputs, hence without the presence of redundant inputs and outputs. 

It is worth mentioning that since $[L]$ is Boolean it will represent an orthogonal basis of the nullspace of the mapping matrix $[B_{M}]$. Hence, the computation of $[L]^{\dag}$ and $([L]^{T})^{\dag}$ will be computationally efficient, because the matrix to be inverted is diagonal \citep{RD_2021},\citep{ML_2007}.

\section{Unconstrained Coupling Form}\label{Unconstrained_Coupling_Form}

As referred in section \ref{Introduction}, when coupling state-space models established in modal domain, they must be previously transformed into coupling form in order to get to a minimal-order coupled state-space model. The same statement is valid for the decoupling operation. Hence, this procedure will equally promote the presence of redundant states on the obtained model representative of the substructure to be identified.

A general state vector transformation into coupling form is given as follows \citep{SJO_20072697}:

\begin{equation}\label{eq:state_vector_transformation}
\{z_{\alpha}(t)\}=[T_{\alpha}]\{q_{\alpha}(t)\}=\left\{\begin{matrix}
\dot{y}_{\alpha}^{J}\\
y_{\alpha}^{J}\\
x_{\alpha}^{I}
\end{matrix}\right\}
\end{equation}

where, $[T]\in \mathbb{R}^{n \times n}$ is the matrix responsible for transforming a state-space model into coupling form, $\left\{ q(t) \right\} \in \mathbb{R}^{n \times 1}$ represents a state vector written in modal domain and $\left\{ z(t) \right\} \in \mathbb{R}^{n \times 1}$ represents a state vector transformed into coupling form. Superscripts $I$ and $J$ denote variables related to internal and interface DOFs, respectively. 

A general construction of the $[T_{\alpha}]$ matrix can be described as follows:

\begin{equation}\label{eq:transf_coupling_formn}
[T_{\alpha}]=\left\{\begin{matrix}
C^{J}_{disp,\alpha}A_{\alpha}\\
C^{J}_{disp,\alpha}\\
N_{\alpha}
\end{matrix}\right\}
\end{equation}

where, the first block row transforms the states into the first derivative of the output DOFs and the second block row transforms the states into the output DOFs \citep{FL_1988}. The last block row represented by $N_{\alpha} \in \mathbb{R}^{n-2n_{J} \times n}$ performs a transformation of the internal states and is included to ensure that the dimension of the state-space matrices are kept.

Conversely to the SSS method presented in \citep{SJO_20072697}, LM-SSS couples all the states of the state-space models to be coupled. Hence, the internal states are not kept uncoupled. Therefore, the computation of $[N_{\alpha}]$ does not need to guarantee that the contribution of the interface inputs to the response of the transformed internal states is null. Consequently, the transformation applied to the internal states can be arbitrary, provided that $[T_{\alpha}]$ is full rank and invertible. The simplest procedure to calculate $[N_{\alpha}]$ in order to fulfill this requirement is given as follows:

\begin{equation}\label{eq:Nullspace_N}
\left[\begin{matrix}
C_{disp,\alpha}^{J}A_{\alpha}\\
C_{disp,\alpha}^{J}
\end{matrix}
\right][N_{\alpha}]^T=0
\end{equation}

Equation \eqref{eq:Nullspace_N} will always guarantee that $[T_{\alpha}]$ is full rank (see equation \eqref{eq:transf_coupling_formn}), if $[C_{disp,\alpha}]$ is full row rank. This equation also concludes the development of a new coupling form specially tailored for LM-SSS. Due to its simplicity and for giving almost completely freedom for the transformation of the internal states, this coupling form will be labelled as Unconstrained Coupling Form (UCF).

When compared with the coupling form developed in \citep{SJO_20072697}, UCF holds the advantage of not relying on the selection of a subspace from a nullspace. The selection of this subspace is difficult and strongly influences the obtained results \citep{mg_2013}. If UCF is compared with the coupling form presented in \textcolor{black}{\citep[section~4.9.4]{AMRDvMTPATMR_2020}} it has the advantage of just requiring the computation of one nullspace.

Note that, before transforming a general state-space model into coupling form, its input and output vectors must be partitioned in terms of interface and internal DOFs, and rearranged as follows.

    \begin{subequations}\label{zero1}
        \noindent
        \begin{tabularx}{\linewidth}{XX}
        \begin{equation}
           \{u_{\alpha}(t)\}= \left\{\begin{matrix}
            u^{J}_{\alpha}(t)\\
            u^{I}_{\alpha}(t)
           \end{matrix}
           \right\}   \label{one}
        \end{equation}
        &
        \begin{equation}
            \{y_{\alpha}(t)\}= \left\{\begin{matrix}
             y^{J}_{\alpha}(t)\\
             y^{I}_{\alpha}(t)
            \end{matrix}
            \right\}  \label{two}
        \end{equation}
    \end{tabularx}
    \end{subequations}

The state-space matrices must be rearranged in accordance with the input and output vectors.

\section{Minimal-order coupled state-space models}\label{Minimal-order coupled state-space models}

Unlike the SSS method proposed in \citep{SJO_20072697}, when coupling state-space models previously transformed into coupling form LM-SSS is not able to directly compute a minimal-order coupled state-space model \citep{RD_2021}. However, since these models were transformed into coupling form, the state vector will contain the interface outputs and respective first derivatives of each substructure, being possible to perform the elimination of the redundant states. Thus, the computation of minimal-order coupled state-space models is still possible.

To better demonstrate how the redundant states must be eliminated, let us assume that the substructures represented in figure \ref{fig:Assembly_rigid_coupling} were coupled by using LM-SSS. Hence, the state vector of the computed coupled state-space model would be given as follows:

\begin{equation}\label{eq:coupled_state_vector}
\begin{gathered}
\{\bar{z}(t)\}=\left\{\begin{matrix}
\dot{\bar{y}}_{\alpha}^{J}(t) &
\bar{y}_{\alpha}^{J}(t) &
\bar{x}_{\alpha}^{I}(t) &
\dot{\bar{y}}_{\beta}^{J}(t) &
\bar{y}_{\beta}^{J}(t) &
\bar{x}_{\beta}^{I}(t)
\end{matrix}\right\}^{T}
\end{gathered}
\end{equation}

During the coupling procedure, compatibility is enforced at the interface of the substructures to be coupled (see section \ref{LM_SSS}), thus we have the following equalities.

    \begin{subequations}\label{zero2}
        \noindent
        \begin{tabularx}{\linewidth}{XX}
        \begin{equation}
\{\bar{y}^{J}\}=\{\bar{y}_{\alpha}^{J}\}=\{\bar{y}_{\beta}^{J}\} \label{five}
        \end{equation}
        &
        \begin{equation}
\{\dot{\bar{y}}^{J}\}=\{\dot{\bar{y}}_{\alpha}^{J}\}=\{\dot{\bar{y}}_{\beta}^{J}\} \label{six}
        \end{equation}
    \end{tabularx}
    \end{subequations}

By observing expressions \eqref{eq:coupled_state_vector}, \eqref{five} and \eqref{six}, we conclude that to obtain the minimal realization of the computed coupled state-space model, $\{\bar{y}_{\alpha}^{J}\}$ or $\{\bar{y}_{\beta}^{J}\}$ must be kept while the other vector must be eliminated from the coupled state-space model. The same is valid for the respective first order derivatives. These state eliminations cannot be performed by simply removing one state of each pair of states that present the same physical meaning. Even though the states to be eliminated are redundant, their contribution for the dynamics of the coupled system is not null. Thus, their impact over the dynamics of the assembled structure must be retrieved by the states that are kept on the model. To properly perform these eliminations is suggested the use of the post-processing procedures outlined in \citep{RD_2021}. 

\color{black}

One of the post-processing procedures proposed in \citep{RD_2021}, perhaps the most straightforward to compute a minimal-order coupled state-space model, relies on the use of a Boolean localization matrix, $[L_{T}]$, which was firstly introduced by Gibanica et al. in \citep{MG_2014}. However, we must alert the readers that this matrix is different from the one used in section \ref{Retaining the unique set of interface DOFs} to retain the unique set of DOFs. Therefore, from now on this matrix will be labelled as state Boolean localization matrix. To construct $[L_{T}]$, we must start by constructing a mapping matrix $[B_{T}]$ (from now on tagged as state mapping matrix to avoid confusion with $[B_{M}]$) as follows:

\begin{equation}\label{eq:mapping}
\begin{gathered}
[B_{T}]\{\begin{matrix}
\bar{x}(t)
\end{matrix}\}=\{0\}
\end{gathered}
\end{equation}

$[L_{T}]$ can now be computed from the nullspace of $[B_{T}]$. Hence, we may establish a relation between the state vector of a computed coupled state-space model and its minimal realization as follows \citep{MG_2014}:

\begin{equation}\label{eq:ss_min_order_num}
\begin{gathered}
\{\begin{matrix}
\bar{x}(t)
\end{matrix}\}=[L_{T}]\{\begin{matrix}
x_{min}(t)
\end{matrix}\}
\end{gathered}
\end{equation}

where, $\{x_{min}(t)\}$ represents the minimal realization of $\{\bar{x}(t)\}$.

By using equation \eqref{eq:ss_min_order_num} and by introducing the Moore-Penrose pseudoinverse of matrix $[L_{T}]$, a minimal-order coupled state-space model can be obtained as given below \citep{RD_2021}.

\begin{equation}\label{eq:accel_ss_model_coup_1}
\begin{gathered}
\{\dot{x}_{min}(t)\}=[L_{T}]^{\dag}[\bar{A}][L_{T}]\{x_{min}(t)\}+[L_{T}]^{\dag}[\bar{B}]\{u(t)\}\\
\{\ddot{y}(t)\}=[\bar{C}][L_{T}]\{x_{min}(t)\}+[\bar{D}]\{u(t)\}
\end{gathered}
\end{equation}

\color{black}
It is worth noticing that the procedures here presented and discussed to eliminate the redundant states originated from coupling are still valid and can be applied in a similar way to eliminate the redundant states originated from a decoupling operation.

\color{black}

\section{Comparison of LM-SSS with others SSS techniques}\label{Comparison of LM-SSS with other SSS techniques}

In this section, we aim at performing a comparison between the LM-SSS method and two of the most popular SSS methods, the one derived by Su and Juang (see \citep{SJ_94}, here also labelled as classical SSS) and the method developed by Sj\"oval and Abrahamsson in \citep{SJO_20072697}. 

To enforce both compatibility and equilibrium conditions, the classical SSS method uses a so-called coupling matrix, here denoted as $[T_{SJ}]$. This matrix considers the interface DOFs of the structures to be coupled. Each row of this matrix corresponds to a coupled DOF, whereas each column is associated to the interface DOFs of the substructures to be coupled. Each row of $[T_{SJ}]$ is created by giving a unitary value to the columns associated with the matching DOFs that are coupled to originate the correspondent coupled DOF, and zeros are assigned to the other columns. From this coupling matrix, relations between the unique set of coupled interface outputs and the interface outputs of the substructures to be coupled can be established (representing the compatibility conditions of our problem). A similar relation representing the equilibrium condition can be established between the unique set of coupled interface inputs and the interface inputs of the substructures to be coupled. Both mentioned coupling conditions are given below.

    \begin{subequations}\label{eq:coupling_conditions_SJ}
        \noindent
        \begin{tabularx}{\linewidth}{XX}
        \begin{equation}
            \{y^{J}(t)\}=[T_{SJ}]^T\{\bar{y}^{J}(t)\}  \label{eq:compatibility_SJ}
        \end{equation}
        &
        \begin{equation}
            \{\bar{u}^{J}(t)\}=[T_{SJ}]\{u^{J}(t)\}  \label{eq:equilibrium_SJ}
        \end{equation}
    \end{tabularx}
    \end{subequations}

To calculate the coupled state-space model by following classical SSS, the state-space matrices of the components to be coupled must be partitioned in terms of internal and interface DOFs. Then, a so-called uncoupled diagonal state-space model can be constructed by concatenating in a diagonal form those partitioned state-space matrices as given below.

\begin{equation}\label{eq:diagonalfirstorderformAB_SJ}
\begin{gathered}
\{\dot{x}(t)\}=[A_{D}]\{x(t)\}+\left[\begin{matrix}
B_{D}^{I} & B_{D}^{J}
\end{matrix}
\right]\left\{\begin{matrix}
u^{I}(t)\\
u^{J}(t)
\end{matrix}
\right\}\\
\left\{\begin{matrix}
y^{I}(t)\\
y^{J}(t)
\end{matrix}
\right\}=\left[\begin{matrix}
C_{D}^{I}\\
C_{D}^{J}
\end{matrix}
\right]\{x(t)\}+\left[\begin{matrix}
D_{D}^{II} & D_{D}^{IJ}\\
D_{D}^{JI} & D_{D}^{JJ}
\end{matrix}
\right]\left\{\begin{matrix}
u^{I}(t)\\
u^{J}(t)
\end{matrix}
\right\}
\end{gathered}
\end{equation}

Expressions in \eqref{eq:MatrixABDiagonal} report the concatenating operation with the $[A]$ matrices of the components to be coupled and with the partitioned state-space matrices associated with the internal DOFs.

\begin{equation}\label{eq:MatrixABDiagonal}
\begin{gathered}
[A_{D}]=\left[
\begin{matrix}
A_{\alpha} &  & \\
 & A_{\beta}  & \\
  & & \ddots
\end{matrix}
\right],\ \
[B_{D}^{I}]=\left[
\begin{matrix}
B_{\alpha}^{I} &  & \\
 & B_{\beta}^{I}  & \\
  & & \ddots
\end{matrix}
\right]\\ \ \ \ 
[C_{D}^{I}]=\left[
\begin{matrix}
C_{\alpha}^{I} &  & \\
 & C_{\beta}^{I}  & \\
  & & \ddots
\end{matrix}
\right],\ \
[D_{D}^{II}]=\left[
\begin{matrix}
D_{\alpha}^{II} &  & \\
 & D_{\beta}^{II}  & \\
  & & \ddots
\end{matrix}
\right]
\end{gathered}
\end{equation}

By enforcing the previously defined coupling conditions on the uncoupled diagonal state-space model (see equation \eqref{eq:diagonalfirstorderformAB_SJ}) and after some mathematical manipulations, one is able to obtain the coupled state-space model as follows: 

\begin{equation}\label{eq:diagonalfirstorderformABcoup}
\begin{gathered}
\{\dot{\bar{x}}(t)\}=[\bar A]\{\bar x(t)\}+\left[\begin{matrix}
\bar B^{I} & \bar B^{J}
\end{matrix}
\right]\left\{\begin{matrix}
\bar{u}^{I}(t)\\
\bar{u}^{J}(t)
\end{matrix}
\right\}\\
\left\{\begin{matrix}
\bar{y}^{I}(t)\\
\bar{y}^{J}(t)
\end{matrix}
\right\}=\left[\begin{matrix}
\bar C^{I}\\
\bar C^{J}
\end{matrix}
\right]\{\bar x(t)\}+\left[\begin{matrix}
\bar D^{II} & \bar D^{IJ}\\
\bar D^{JI} & \bar D^{JJ}
\end{matrix}
\right]\left\{\begin{matrix}
\bar{u}^{I}(t)\\
\bar{u}^{J}(t)
\end{matrix}
\right\}
\end{gathered}
\end{equation}

where,

\begin{equation}\label{eq:coupledAmatrix}
\begin{gathered}
[\bar A]=A_{D}+B_{D}^{J}QC_{D}^{J}\\
[\bar B]=\left[\begin{matrix}
B_{D}^{I}+B_{D}^{J}QD_{D}^{JI} & B_{D}^{J}(D_{D}^{JJ})^{-1}T_{SJ}^{T}S^{-1}
\end{matrix}
\right]\\
[\bar C]=\left[\begin{matrix}
C_{D}^{I}+D_{D}^{IJ}QC_{D}^{J}\\
S^{-1}T_{SJ}(D_{D}^{JJ})^{-1}C_{D}^{J}
\end{matrix}
\right]\\
[\bar D]=\left[\begin{matrix}
D_{D}^{II}+D_{D}^{IJ}QD_{D}^{JI} & D_{D}^{IJ}(D_{D}^{JJ})^{-1}T_{SJ}^{T} S^{-1}\\
S^{-1}T_{SJ}(D_{D}^{JJ})^{-1}D_{D}^{JI} & S^{-1}
\end{matrix}
\right]
\end{gathered}
\end{equation}

where the auxiliary variables $[S]$ and $[Q]$ are given below.

\begin{equation}\label{eq:variableS}
[S]=T_{SJ}(D_{D}^{JJ})^{-1}T_{SJ}^{T}
\end{equation}

\begin{equation}\label{eq:variableQ}
[Q]=(D_{D}^{JJ})^{-1}T_{SJ}^{T}S^{-1}T_{SJ}(D_{D}^{JJ})^{-1}-(D_{D}^{JJ})^{-1}
\end{equation}

Since the compatibility and equilibrium conditions establish direct relations between the unique set of interface outputs and inputs and the interface outputs and inputs of the substructures to be coupled (see equations \eqref{eq:compatibility_SJ} and \eqref{eq:equilibrium_SJ}), a state-space model without the presence of redundant DOFs is directly obtained. This fact represents one of the major differences between LM-SSS and classical SSS. Besides the pointed difference, we may also claim that the implementation of LM-SSS is simpler, because by using LM-SSS no partition of the state-space matrices is needed to compute the diagonal coupled state-space model used to calculate the coupled state-space model (see section \ref{LM_SSS}). Moreover, classical SSS requires the performance of two matrix inversions, while LM-SSS requires only one, this might be a major advantage when dealing with real-time substructuring, specially when dealing with systems presenting time varying performances and composed by many interface DOFs. In these cases, state-space models for each variation on the dynamic properties of the components to be coupled would have to be computed and then coupled. Hence, several coupling operations would be necessary. For this reason, the possibility of performing just one matrix inversion instead of two for each coupling operation could represent a dramatic decrease of the computational cost associated with the real-time substructuring application. Furthermore, the matrix to be inverted by coupling with LM-SSS is of dimension $n_{J}\times n_{J}$ (see expression \eqref{eq:coupled_matrices_value}) (where, $n_{J}$ represents the number of unique interface DOFs of the coupled structure), while the matrices to be inverted during a coupling operation by using classical SSS are of dimension $n_{J}\times n_{J}$ and $2n_{J} \times 2n_{J}$ (see expression \eqref{eq:coupledAmatrix}). Thus, as the number of interface DOFs to be coupled increases, more dramatic would be the decrease on the computational cost by using the LM-SSS technique instead of the classical SSS method. 

To analyze the method developed by Sj\"oval and Abrahamsson in \citep{SJO_20072697}, we will assume that it is intended to perform the coupling of two substructures tagged as $\alpha$ and $\beta$ to obtain the correspondent displacement coupled state-space model. The method developed in \citep{SJO_20072697} requires the previous transformation of the state-space models to be coupled into the coupling form proposed in the same reference (from now on labelled as SACF, standing for Sj\"ovall and Abrahamsson Coupling Form). By transforming a state-space model into $SACF$, its state-space matrices are transformed into a particular structure as follows

\begin{equation}\label{eq:MatrixABsubstructure1}
\begin{gathered}
\left\{\begin{matrix}
\ddot{y}^{J}_{i}(t)\\
\dot{y}^{J}_{i}(t)\\
\dot{x}^{I}_{i}(t)
\end{matrix}
\right\}=\left[
\begin{matrix}
A_{i}^{vv} & A_{i}^{vd} & A_{i}^{vI} \\
I & 0 & 0 \\
0 & A_{i}^{Id}  & A_{i}^{II}
\end{matrix}
\right]\left\{\begin{matrix}
\dot{y}^{J}_{i}(t)\\
y^{J}_{i}(t)\\
x^{I}_{i}(t)
\end{matrix}
\right\}+\left[
\begin{matrix}
B_{i}^{vv} & B_{i}^{vI}\\
0 & 0 \\
0 & B_{i}^{II}
\end{matrix}
\right]\left\{\begin{matrix}
\dot{u}^{J}_{i}(t)\\
u^{I}_{i}(t)
\end{matrix}
\right\}\\
\left\{\begin{matrix}
y^{J}_{i}(t)\\
y^{I}_{i}(t)
\end{matrix}
\right\}=\left[
\begin{matrix}
0 & I & 0\\
C_{i}^{Iv} & C_{i}^{Id}  & C_{i}^{II}
\end{matrix}
\right]\left\{\begin{matrix}
\dot{y}^{J}_{i}(t)\\
y^{J}_{i}(t)\\
x^{I}_{i}(t)
\end{matrix}
\right\}
\end{gathered}
\end{equation}

where, subscript $i$ denotes vectores/matrices associated with structure $i$, in this case $i$ may either be $\alpha$ or $\beta$.

After transforming the state-space models to be coupled into coupling form, both compatibility and equilibrium conditions must be established, as given below.

    \begin{subequations}\label{eq:coupling_conditions_Sjoval_Abrahamsson}
        \noindent
        \begin{tabularx}{\linewidth}{XX}
        \begin{equation}
        \left\{\begin{matrix}
                y^{J}_{\alpha}(t)\\
                y^{J}_{\beta}(t)
                \end{matrix}\right\}=\left[\begin{matrix}
                I\\
                I
                \end{matrix}\right]\{\bar{y}^{J}(t)\}  \label{eq:compatibility_Sjoval_Abrahamsson}
        \end{equation}
        &
        \begin{equation}
            \{\bar{u}^{J}(t)\}=\left[\begin{matrix}
                I & I
                \end{matrix}\right]\left\{\begin{matrix}
                u_{\alpha}^{J}(t)\\
                u_{\beta}^{J}(t)
                \end{matrix}\right\}  \label{eq:equilibrium_Sjoval_Abrahamsson}
        \end{equation}
    \end{tabularx}
    \end{subequations}

In the previous expressions, $[I]$ is an identity matrix of dimension $n_{J}\times n_{J}$.

To compute the coupled state-space model, one must start by adding the first block row of the state equations of the state-space models representative of the dynamics of substructures $\alpha$ and $\beta$. Then, by imposing both compatibility and equilibrium conditions (see equations \eqref{eq:compatibility_Sjoval_Abrahamsson} and \eqref{eq:equilibrium_Sjoval_Abrahamsson}), the first block row of the coupled state-space model is computed (see \citep{AL_2012}). Lastly, by correctly placing the remaining state-space matrix elements of the state-space models to be coupled, the coupled state-space model can be obtained as follows  

\begin{equation}\label{eq:Coupled_SSM_Sjoval_Abrahamsson}
\begin{gathered}
\left\{\begin{matrix}
\ddot{\bar{y}}^{J}(t)\\
\dot{\bar{y}}^{J}(t)\\
\dot{x}^{I}_{\alpha}(t)\\
\dot{x}^{I}_{\beta}(t)
\end{matrix}
\right\}=\left[
\begin{matrix}
\bar{A}^{vv} & \bar{A}^{vd} & \bar{A}_{\alpha}^{vI} & \bar{A}_{\beta}^{vI}\\
I & 0 & 0 & 0 \\
0 & A_{\alpha}^{Id} & A_{\alpha}^{II} & 0\\
0 & A_{\beta}^{Id} & 0 & A_{\beta}^{II}
\end{matrix}
\right]\left\{\begin{matrix}
\dot{\bar{y}}^{J}(t)\\
\bar{y}^{J}(t)\\
x^{I}_{\alpha}(t)\\
x^{I}_{\beta}(t)
\end{matrix}
\right\}+\left[
\begin{matrix}
\bar{B}^{vv} & \bar{B}_{\alpha}^{vI} & B_{\beta}^{vI}\\
0 & 0 & 0\\
0 & B_{\alpha}^{II} & 0\\
0 & 0 & B_{\beta}^{II}\\
\end{matrix}
\right]\left\{\begin{matrix}
\bar{u}^{J}(t)\\
u^{I}_{\alpha}(t)\\
u^{I}_{\beta}(t)
\end{matrix}
\right\}\\
\left\{\begin{matrix}
\bar{y}^{J}(t)\\
y^{I}_{\alpha}(t)\\
y^{I}_{\beta}(t)
\end{matrix}
\right\}=\left[
\begin{matrix}
0 & I & 0 & 0\\
C_{\alpha}^{Iv} & C_{\alpha}^{Id}  & C_{\alpha}^{II} & 0\\
C_{\beta}^{Iv} & C_{\beta}^{Id} & 0 & C_{\beta}^{II}
\end{matrix}
\right]\left\{\begin{matrix}
\dot{\bar{y}}^{J}(t)\\
\bar{y}^{J}(t)\\
x^{I}_{\alpha}(t)\\
x^{I}_{\beta}(t)
\end{matrix}
\right\}
\end{gathered}
\end{equation}

where,

\begin{equation}
\bar{A}^{vv}=B_{\alpha}^{vv}\Gamma A_{\beta}^{vv}+B_{\beta}^{vv}\Gamma A_{\alpha}^{vv}
\end{equation}
\begin{equation}
\bar{A}^{vd}=B_{\alpha}^{vv}\Gamma A_{\beta}^{vd}+B_{\beta}^{vv}\Gamma A_{\alpha}^{vd}
\end{equation}
\begin{equation}
\bar{A}_{\alpha}^{vI}=B_{\beta}^{vv}\Gamma A_{\alpha}^{vI}
\end{equation}
\begin{equation}
\bar{A}_{\beta}^{vI}=B_{\alpha}^{vv}\Gamma A_{\beta}^{vI}
\end{equation}
\begin{equation}
\bar{B}^{vv}=B_{\alpha}^{vv}\Gamma B_{\beta}^{vv}
\end{equation}
\begin{equation}
\bar{B}_{\alpha}^{vI}=B_{\beta}^{vv}\Gamma B_{\alpha}^{vI}
\end{equation}  
\begin{equation}
\bar{B}_{\beta}^{vI}=B_{\alpha}^{vv}\Gamma B_{\beta}^{vI}
\end{equation}

with $\Gamma=(B_{\alpha}^{vv}+B_{\beta}^{vv})^{-1}$.

The coupled state-space model given by expression \eqref{eq:Coupled_SSM_Sjoval_Abrahamsson} is already obtained in a minimal-order form, not presenting redundant states originated from the coupling operation. Furthermore, it does not present any redundant DOF, since the coupling conditions were directly established between the unique set of interface outputs and inputs and the outputs and inputs of the components to be coupled (see equations \eqref{eq:compatibility_Sjoval_Abrahamsson} and \eqref{eq:equilibrium_Sjoval_Abrahamsson}). Thus, the post-processing procedures needed to retain the unique set of DOFs (see section \ref{Retaining the unique set of interface DOFs}) and to compute a minimal-order coupled state-space model (see section \ref{Minimal-order coupled state-space models}) when coupling with LM-SSS are not required when using the technique developed by Sj\"oval and Abrahamsson. Nevertheless, it is worth mentioning that LM-SSS enables the simultaneous coupling of an unlimited number of substructures, while the technique suggested by Sj\"oval and Abrahamsson can do it on just two substructures at same time. Moreover, it is worth mentioning that coupling with Sj\"oval and Abrahamsson approach is possible, because by transforming a state-space model into SACF, one makes sure that the contribution of the interface inputs to the response of the internal states is null. If this condition does not hold, one could not directly insert state-space matrix elements of the models to be coupled into the coupled state-space model (see expression \eqref{eq:Coupled_SSM_Sjoval_Abrahamsson}). By using LM-SSS the internal states are included in the coupling operation, opening the possibility of applying a arbitrary transformation to the internal states, provided that the transformation matrix is full rank and invertible. For this reason, LM-SSS does not require the previous transformation of the state-space models to be coupled into coupling form, unless a minimal-order coupled state-space model is intended to be computed. Even if we are interested in computing a minimal-order coupled state-space model with LM-SSS, we have the possibility of coupling state-space models previously transformed into $UCF$ instead of $SACF$. The possibility of use $UCF$ is an important advantage, because it avoids the need of selecting a subspace from a nullspace as required by $SACF$. The selection of the mentioned subspace is difficult and strongly influence the accuracy of transforming a state-space model into $SACF$ \citep{mg_2013}.

\color{black}

\section{Numerical Example}\label{Numerical example}

\begin{figure}[ht]
\centering
    \includegraphics[width=0.6\textwidth]{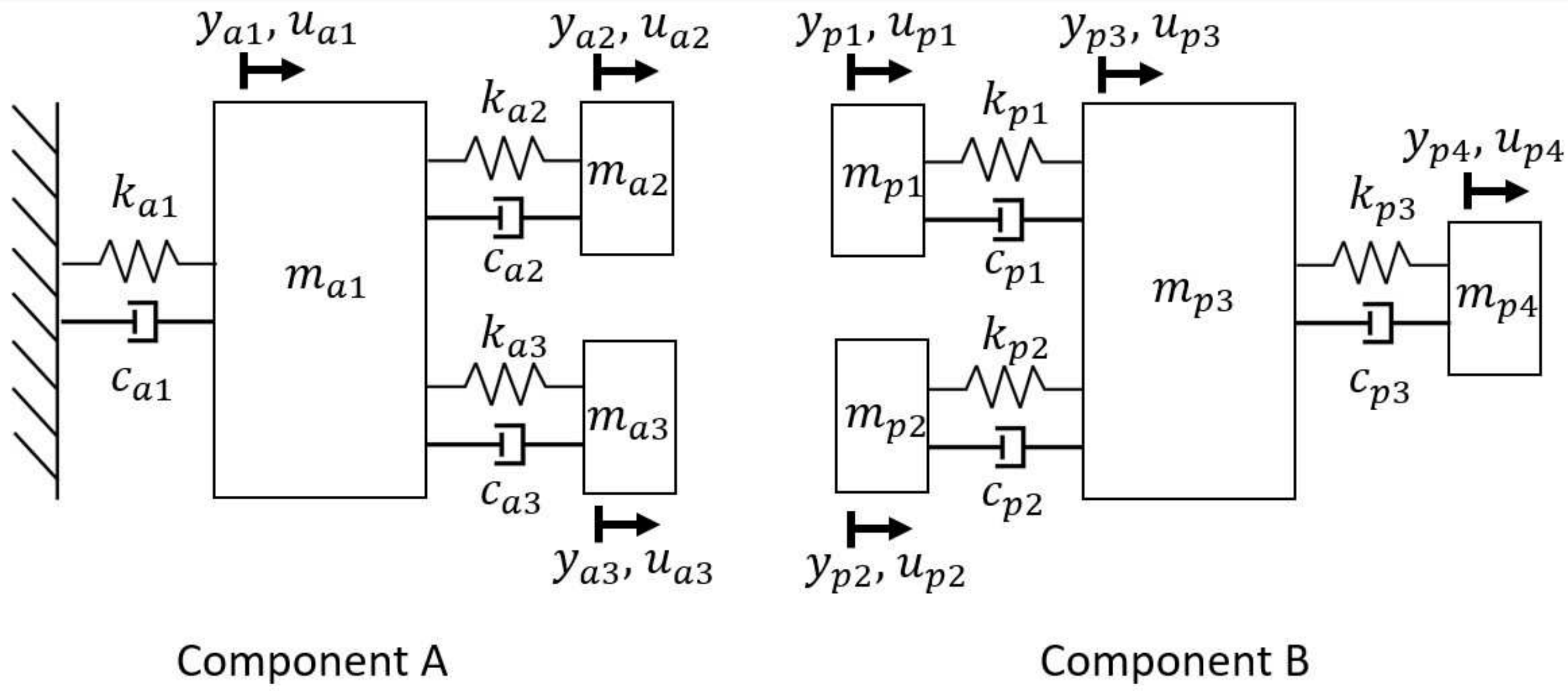}
    \caption{Uncoupled components.}
     \label{fig:uncoupled_components_A_B}
\end{figure}

\begin{figure}[ht]
\centering
    \includegraphics[width=0.6\textwidth]{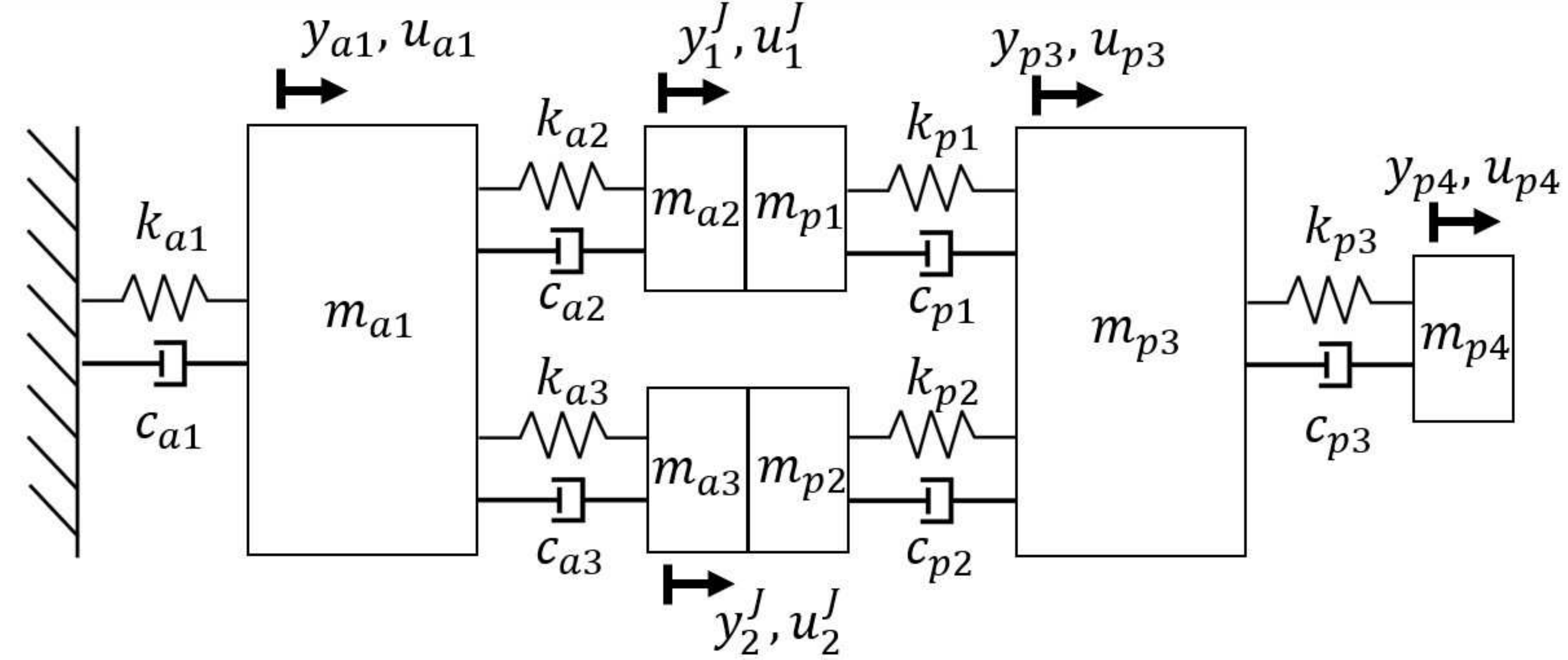}
    \caption{Assembled system.}
     \label{fig:Assembled_system_AB}
\end{figure}

The approaches discussed in the previous sections will be demonstrated hereafter by adopting a numerical example. The dynamic system synthesized is represented by the two components shown in figure \ref{fig:uncoupled_components_A_B}. The assembled structure obtained when both components are coupled is presented in figure \ref{fig:Assembled_system_AB}, being the values of the physical parameters indicated in both figures given in table \ref{table:T1.1}.

\setlength{\tabcolsep}{20pt}

\begin{table}[ht]
\centering
\caption{Physical parameter values}
\begin{tabular}{@{}llll@{}}
\toprule
$i$ \ & $m_{i}$ ($kg$) \ & $c_{i}$ ($Nsm^{-1}$) \ & $k_{i}$ ($Nm^{-1}$)  \\ \midrule
$a1$ & 10 & 30             & $1.5\times 10^{5}$\\
$a2$ & 3 & 50             & $5\times 10^{5}$\\
$a3$ & 3 & 50             & $4.5\times 10^{5}$\\
$p1$ & 5 & 50             & $1\times 10^{5}$\\
$p2$ & 7 & 50             & $1.5\times 10^{5}$\\
$p3$ & 10 & 10             & $5\times 10^{3}$\\
$p4$ & 1 & -             & -\\ \midrule  
\label{table:T1.1}
\end{tabular}
\end{table}

Firstly, the estimation of state-space models from the FRFs of components $A$ and $B$ is performed. Then, to validate UCF presented in section \ref{Unconstrained_Coupling_Form}, the identified state-space models will be transformed into this coupling form and into the ones presented in \citep{SJO_20072697} and \textcolor{black}{\citep[section~4.9.4]{AMRDvMTPATMR_2020}}. The FRFs of the untransformed and of the three transformed identified models of each component are then compared (section \ref{Identified State-Space Models}). 

In section \ref{Coupling Results}, the coupling of the identified state-space models will be performed by using the LM-SSS method (see section \ref{LM_SSS}), being the FRFs of the coupled model compared with the ones of the exact model and with the coupled FRFs obtained by applying LM FBS \citep{DK06} to couple the FRFs of the identified models. To validate LM-SSS to compute minimal-order models, this SSS technique will be applied to compute coupled models from the identified models transformed into different coupling forms, being the elimination of the redundant states present on those coupled models performed by following the procedures described in section \ref{Minimal-order coupled state-space models}. The FRFs of the computed minimal-order models will be, then, compared with the ones of the coupled model computed from the untransformed identified models.

Finally, in section \ref{Decoupling Results} LM-SSS will be evaluated to perform decoupling in order to identify the state-space model of the component $B$. Then, the FRFs of the obtained model will be compared with the FRFs of the exact model and the FRFs of the model identified by using the system identification method and with the FRFs obtained by using LM FBS decoupling. To demonstrate that the procedures described in section \ref{Minimal-order coupled state-space models} are valid to eliminate redundant states originated from a decoupling operation, the identification of the state-space model of component $B$ will be repeated by using state-space models previously transformed into coupling form. The FRFs of the obtained models are then compared with the FRFs of the model obtained from LM-SSS decoupling by using untransformed state-space models.

\subsection{Identified State-Space Models}\label{Identified State-Space Models}

The computation of the exact state-space model of the components and of the assembled system was performed by using its mass, stiffness and damping matrices in accordance with equations \eqref{eq:accel_ss_model} and \eqref{eq:MatrixABCD_numerical}. The mass and stiffness matrices were established by using the Lagrange equations \citep{HRH_1997}, while for simplicity the damping matrix of each substructure was constructed from the respective stiffness one by replacing the stiffness terms with the damping ones.

To approximate the numerical example under study to an experimental one, artificial noise was introduced into the FRFs of both components. The computation of the perturbed FRFs was performed by following the procedure used in \citep{SJO_20072697}. This procedure perturbs the real and imaginary parts of each element of the FRFs according to equation \eqref{eq:perturbing_FRFs}.

\begin{equation}\label{eq:perturbing_FRFs}
H_{S,ij}(\omega_{k})=H_{S,ij}(\omega_{k})+\gamma_{ijk}+j\theta_{ijk}
\end{equation}

where, subscript $S$ denotes the substructure to which the FRFs that are being perturbed belong, while subscripts $i$, $j$ and $k$ denote the output, input and the discrete frequency of the FRF term that is being perturbed. Variables $\gamma$ and $\theta$ are Gaussian distributed independent stochastic variables with zero mean and a standard deviation, for the specific case, assumed to be equal to $5 \times 10^{-3}$ $ms^{-2}N^{-1}$. \textcolor{black}{Note that, by perturbing the FRFs in accordance with equation \eqref{eq:perturbing_FRFs}, noise is added only to the response (output) part of the FRFs.}

\begin{figure}[ht]
\centering
    \includegraphics[width=1\textwidth]{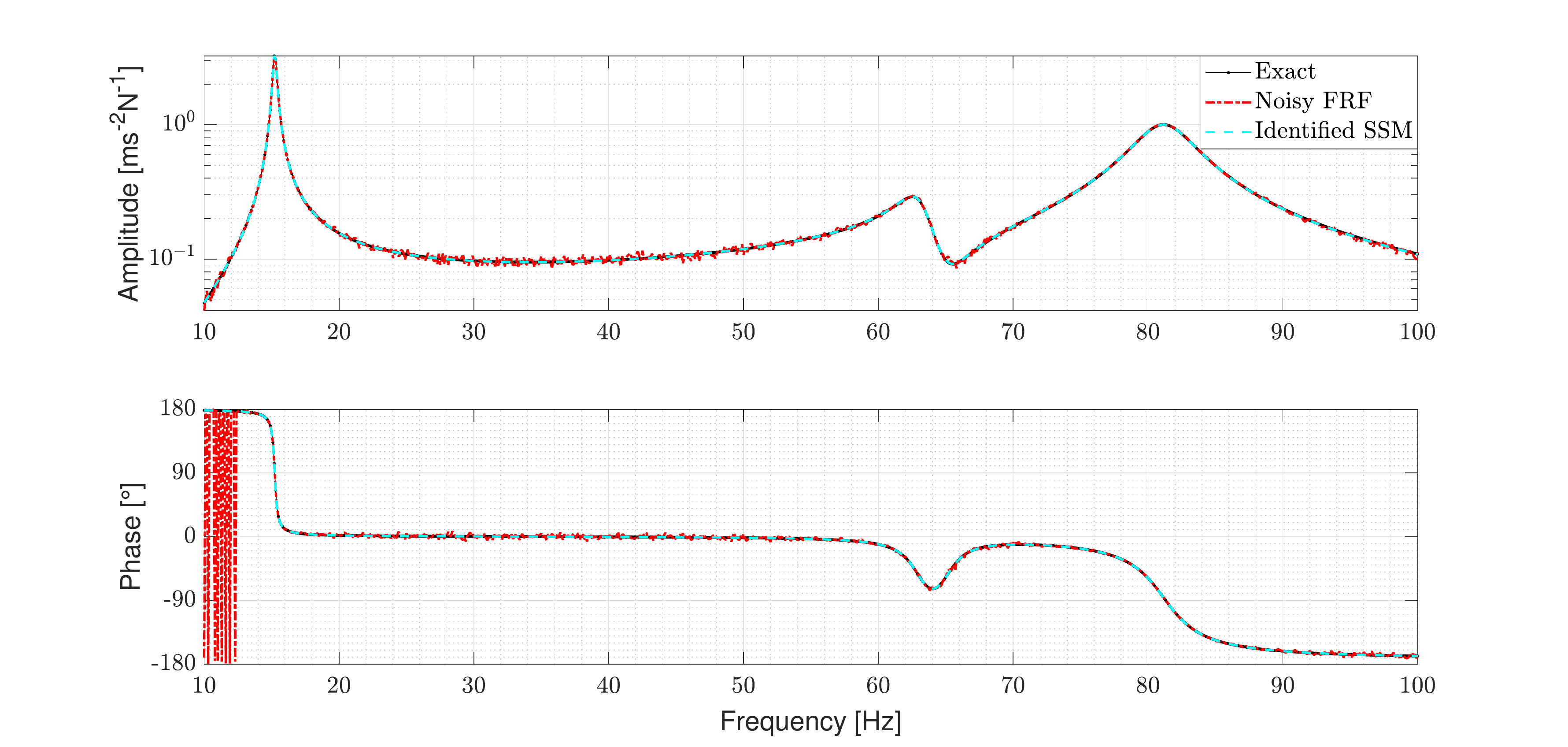}
    \caption{Accelerance FRF of the component $A$, whose output is the DOF $a3$ and the input is the DOF $a1$.}
     \label{fig:component_A_identified}
\end{figure}

\begin{figure}[ht]
\centering
    \includegraphics[width=1\textwidth]{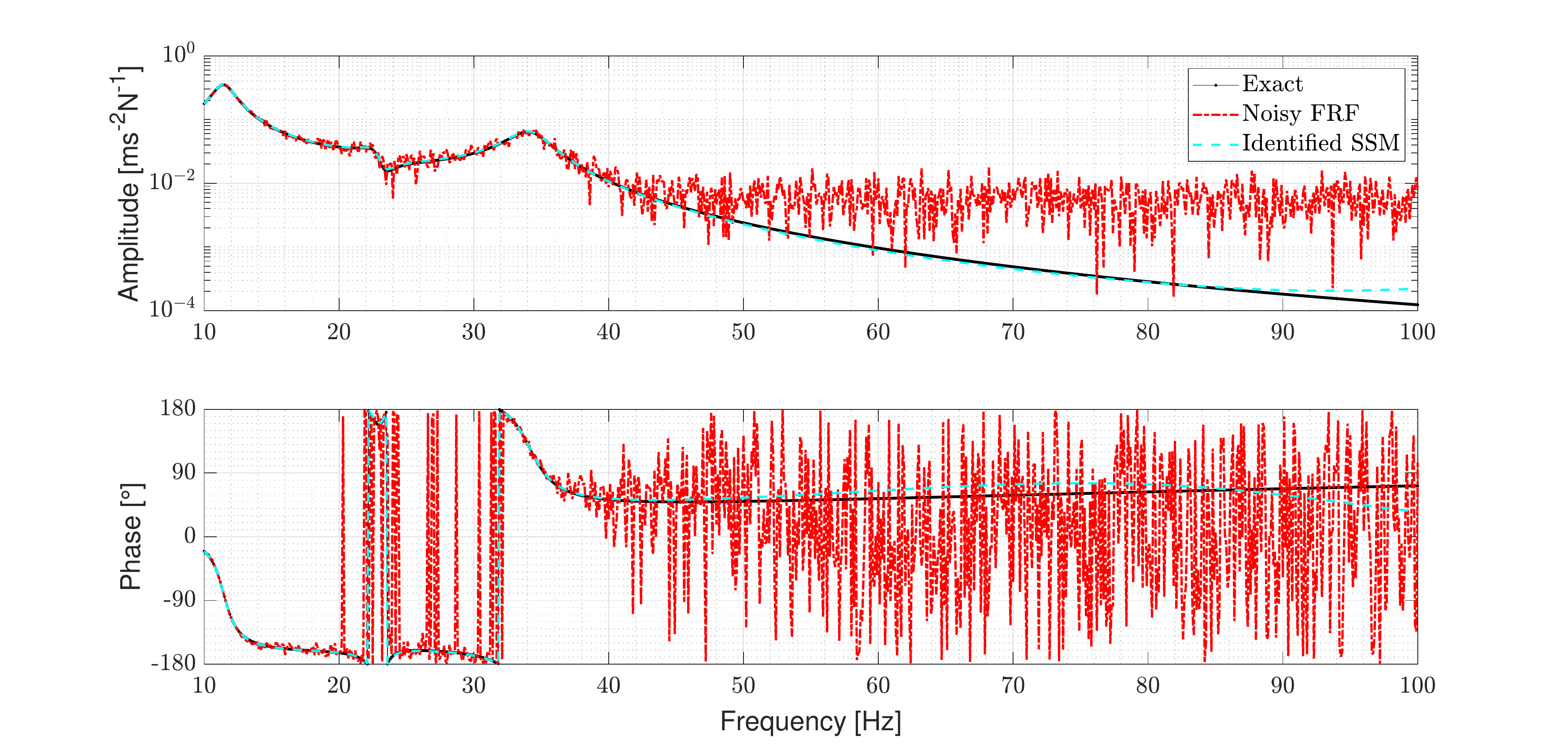}
    \caption{Accelerance FRF of the component $B$, whose output is the DOF $p4$ and the input is the DOF $p1$.}
     \label{fig:component_B_identified}
\end{figure}

To identify state-space models from the computed noisy FRFs, \textcolor{black}{PolyMAX} \citep{BPet_2004395} and the Maximum Likelihood Modal Parameter method (ML-MM) \citep{MEL_2015567} were applied to estimate the modal parameters from the noisy FRFs. \textcolor{black}{It is worth mentioning that the Simcenter Testlab\textsuperscript{\textregistered} implementation of the approaches was exploited.} Then, these identified parameters were used to construct the respective state-space models. The comparison of a noisy FRF with the FRF of the exact model and the FRF of the identified model for component $A$ is given in figure \ref{fig:component_A_identified}, while the same comparison for component $B$ is shown in figure \ref{fig:component_B_identified}.

\begin{figure}[ht]
\centering
    \includegraphics[width=1\textwidth]{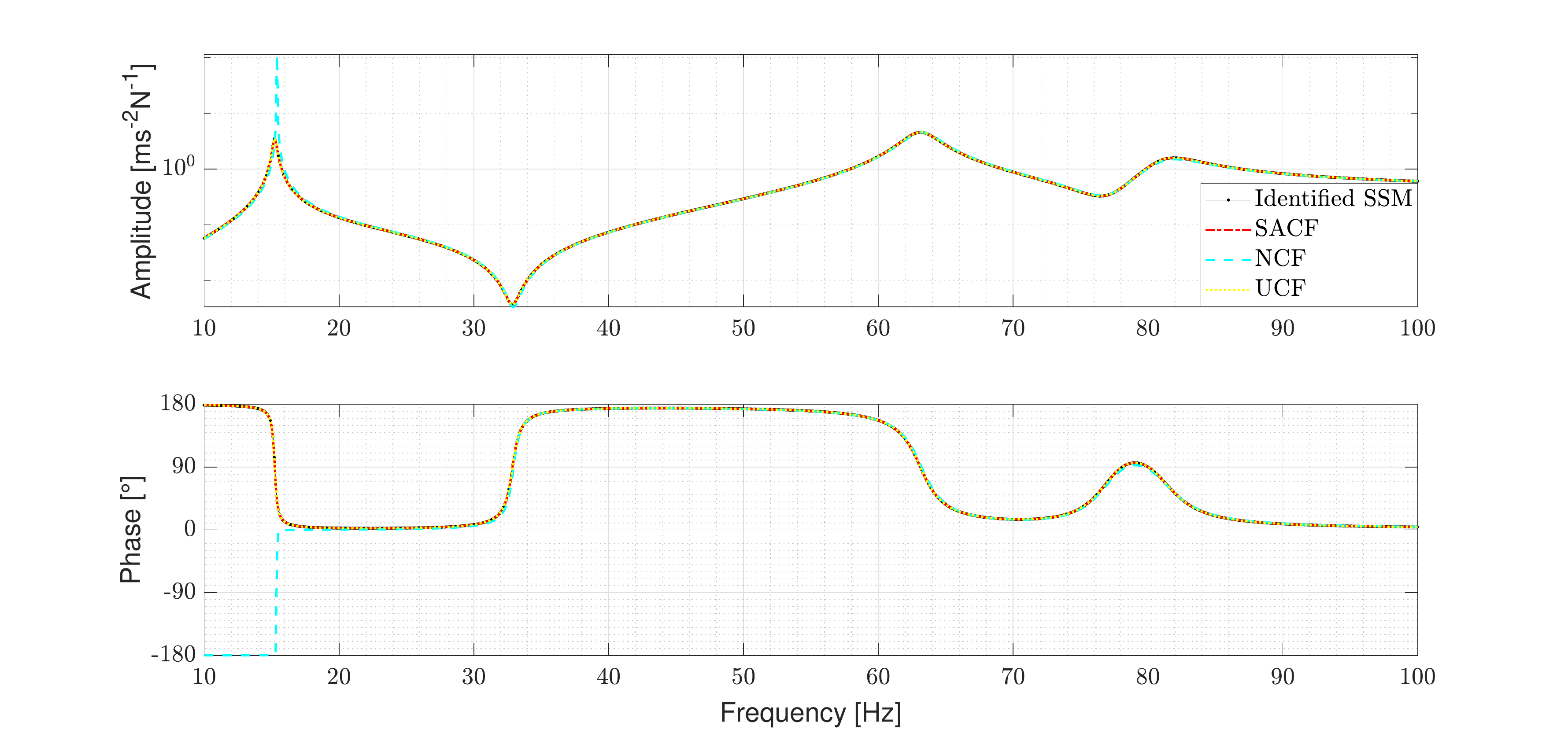}
    \caption{Accelerance FRF of the component $A$, whose output is the DOF $a3$ and the input is the DOF $a3$.}
     \label{fig:component_A_identified_coupling_form}
\end{figure}

\begin{figure}[ht]
\centering
    \includegraphics[width=1\textwidth]{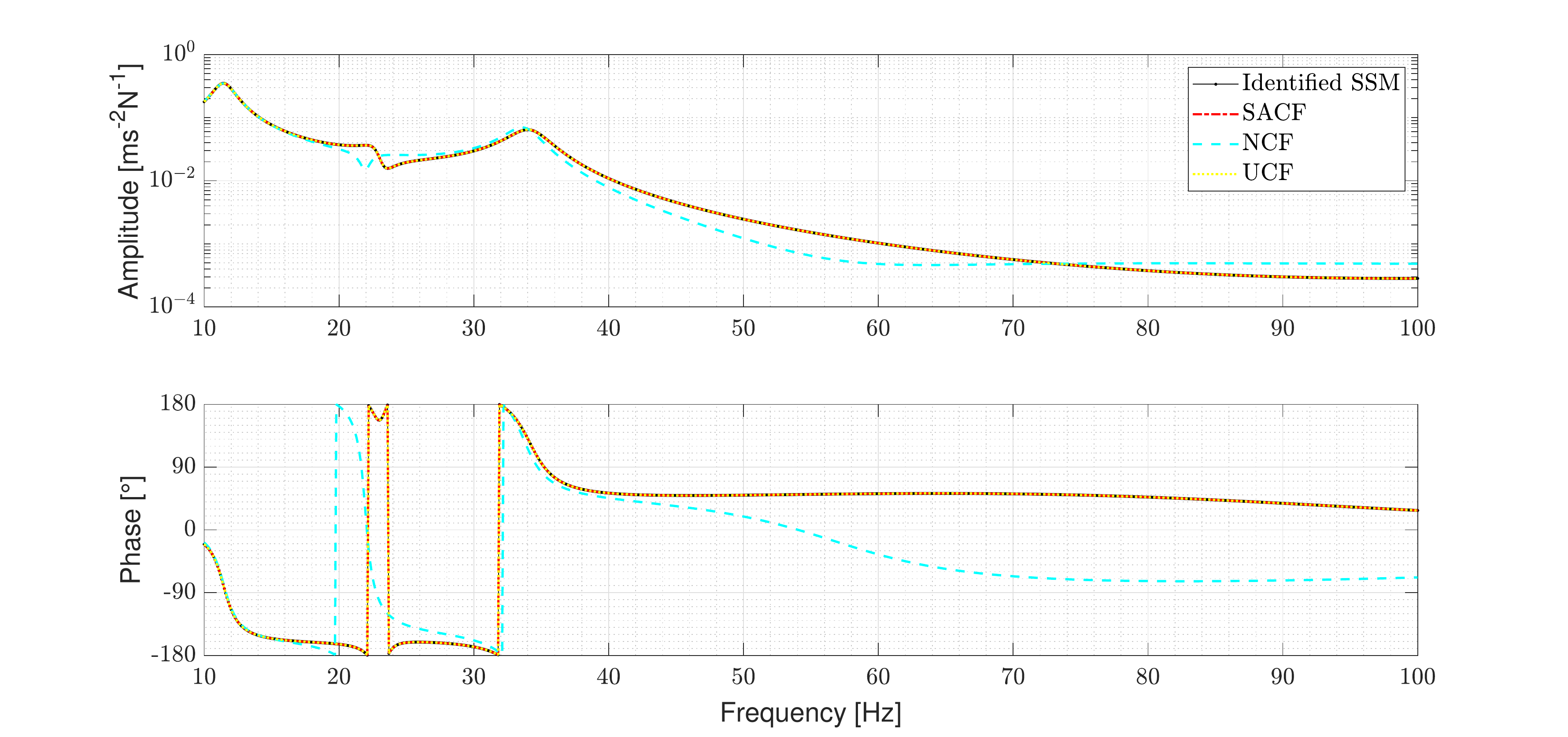}
    \caption{Accelerance FRF of the component $B$, whose output is the DOF $p1$ and the input is the DOF $p4$.}
     \label{fig:component_B_identified_coupling_form}
\end{figure}

Afterwards, the identified state-space models were transformed into the coupling forms presented in \textcolor{black}{\citep{SJO_20072697} (here labelled as $SACF$) and \textcolor{black}{\citep[section~4.9.4]{AMRDvMTPATMR_2020}} (from now on labelled as NCF, standing for New Coupling Form)} and into UCF presented in section \ref{Unconstrained_Coupling_Form}. In figures \ref{fig:component_A_identified_coupling_form} and \ref{fig:component_B_identified_coupling_form} the comparison of one FRF of the identified untransformed model and of the same model transformed into different coupling forms is shown for components $A$ and $B$ respectively.

By analyzing figures \ref{fig:component_A_identified} and \ref{fig:component_B_identified} it is evident that the system identification method used was able to filter-out the artificial noise introduced on the FRFs of both components, thus leading to an accurate identification of the state-space models of both components (the identified models of components $A$ and $B$ were composed by $n_{A}=18$ and $n_{B}=22$ states, respectively). However, it was found that the state-space models were not passive. Looking at figures \ref{fig:component_A_identified_coupling_form} and \ref{fig:component_B_identified_coupling_form}, it is well evident that for both the components the FRF of the identified transformed models into \textcolor{black}{SACF} and UCF is very well matching the one of the untransformed model. Hence, we may conclude that the transformation of both identified state-space models into these two different coupling forms was successfully performed. However, it was found that the transformation into \textcolor{black}{NCF} lead to poor results. This poor performance might be explained by its transformation matrix being close to be singular. Indeed, as proven in \ref{Appendix}, we have no guarantee that the \textcolor{black}{NCF} transformation matrix will be always well-conditioned and full rank. For the other coupling forms this is not verified. Provided that the input and output state-space matrices of the state-space model to be transformed are full column rank and full row rank, respectively \citep{MS_2015}, by using \textcolor{black}{SACF} we have freedom to select a subspace from the computed nullspace that makes the computation of a full rank transformation matrix possible \citep{SJO_20072697}. Whereas, by using UCF, if the output state-space matrix of the state-space model to be transformed is full row rank, the transformation matrix will always be full rank as presented in section \ref{Unconstrained_Coupling_Form}.

Note that, due to the poor performance of \textcolor{black}{NCF} transformation, from now on the performance of UCF will be just compared with the one of \textcolor{black}{SACF}.

\subsection{Coupling Results}\label{Coupling Results}

\begin{figure}[ht]
\centering
    \includegraphics[width=1 \textwidth]{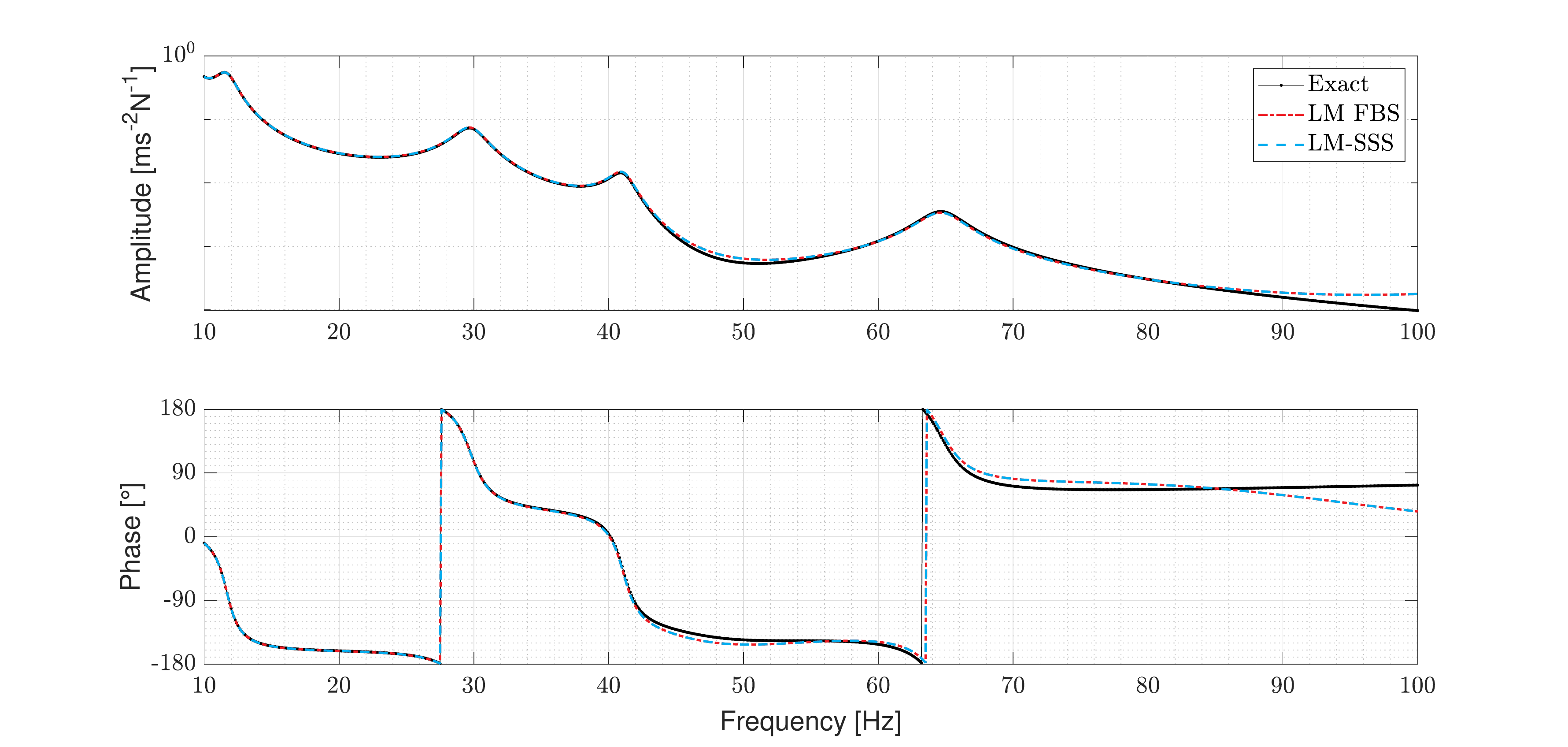}
    \caption{Accelerance FRF of the coupled system, whose output is the \textcolor{black}{internal} DOF $p4$ and the input is the interface DOF $1$.}
     \label{fig:Assembled_system_AB_coupled}
\end{figure}

In a first instance, the LM-SSS method was used to couple the untransformed identified state-space models. The obtained model presented $n_{A}+n_{B}=40$ states as expected. To compute this coupled model, the retention of the unique set of DOFs was performed by following the procedure outlined in section \ref{Retaining the unique set of interface DOFs}. Then, the coupling was made by using LM FBS to couple the FRFs of the identified state-space models. Figure \ref{fig:Assembled_system_AB_coupled} shows the comparison of a coupled FRF obtained by using LM FBS method with the same FRF of the exact assembled state-space model and of the non minimal-order coupled model. 

By analyzing figure \ref{fig:Assembled_system_AB_coupled}, it is evident that the FRF of the exact assembled model and the FRF of the non minimal-order coupled model are well matching. Furthermore, the FRF computed from LM FBS perfectly matches the one of the coupled state-space model. However, the coupled state-space model was found to be unstable, presenting five real positive poles. The presence of those non-physical poles might be a consequence of the non-passivity of the identified state-space models \citep{SJO_20072697}.

\begin{figure}[ht]
\centering
    \includegraphics[width=1\textwidth]{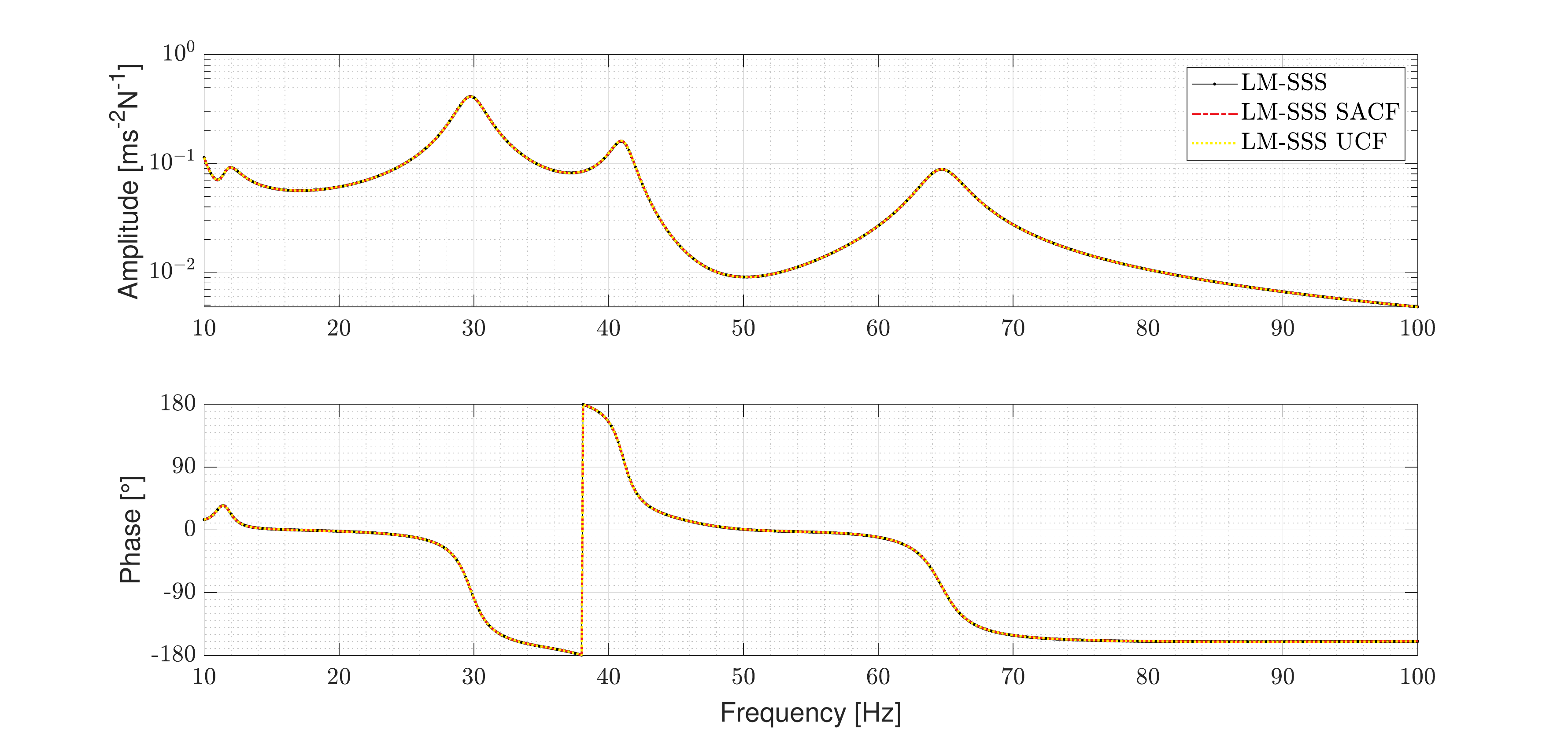}
    \caption{Accelerance FRF of the coupled system, whose output is the interface DOF $1$ and the input is the DOF $p3$.}
     \label{fig:Assembled_system_AB_coupling_form}
\end{figure}

To evaluate the performance of LM-SSS to compute minimal-order models by coupling state-space models previously transformed into coupling form and by using the procedures described in section \ref{Minimal-order coupled state-space models}, LM-SSS was applied to compute minimal-order coupled models (it was found that these models presented $n_{A}+n_{B}-2n_{J}=36$ states as expected) from the identified models transformed into two different coupling forms (\textcolor{black}{SACF} and UCF). For ease of implementation, the retention of the unique set of DOFs of the computed coupled state-space models was performed by following the procedure outlined in section \ref{Retaining the unique set of interface DOFs}. For the same reason, the post-processing procedure presented in \citep{RD_2021} that relies on the use of a \textcolor{black}{state} Boolean localization matrix $[L_{T}]$ was used to eliminate the redundant states. Figure \ref{fig:Assembled_system_AB_coupling_form} shows the comparison between the FRF of the non-minimal order coupled model and the \textcolor{black}{two} FRFs obtained by the different minimal order models. 

Figure \ref{fig:Assembled_system_AB_coupling_form} proves that LM-SSS and the procedures described in section \ref{Minimal-order coupled state-space models} to obtain minimal order coupled models are consistent.  For the two different minimal-order coupled state-space models were found three non-physical poles (i.e. real positive poles), which might be a direct consequence of the transformed identified state-space models not be passive, as pointed out in \citep{SJO_20072697}. The decrease of the number of non-physical poles, when compared with the non-minimal order coupled model is a consequence of the elimination of the redundant states. This verification highlights the importance of removing these spurious states from the coupled state-space models. \color{black}{Nevertheless, we must highlight that even by using LM-SSS method, the computed coupled state-space models might be unstable, specially if the state-space models involved on the coupling operation are not passive, as reported by \citep{SJO_20072697}.}
\color{black}

\subsection{Decoupling Results}\label{Decoupling Results}

To validate LM-SSS to perform decoupling, this method was applied to decouple the identified state-space model of component $A$ (see section \ref{Identified State-Space Models}) from the non minimal-order coupled state-space model computed in section \ref{Coupling Results}. The same procedure was performed by decoupling from the FRFs of the non minimal-order coupled model the identified FRFs of component $A$ by using the LM FBS technique. 

\begin{figure}[ht]
\centering
    \includegraphics[width=1\textwidth]{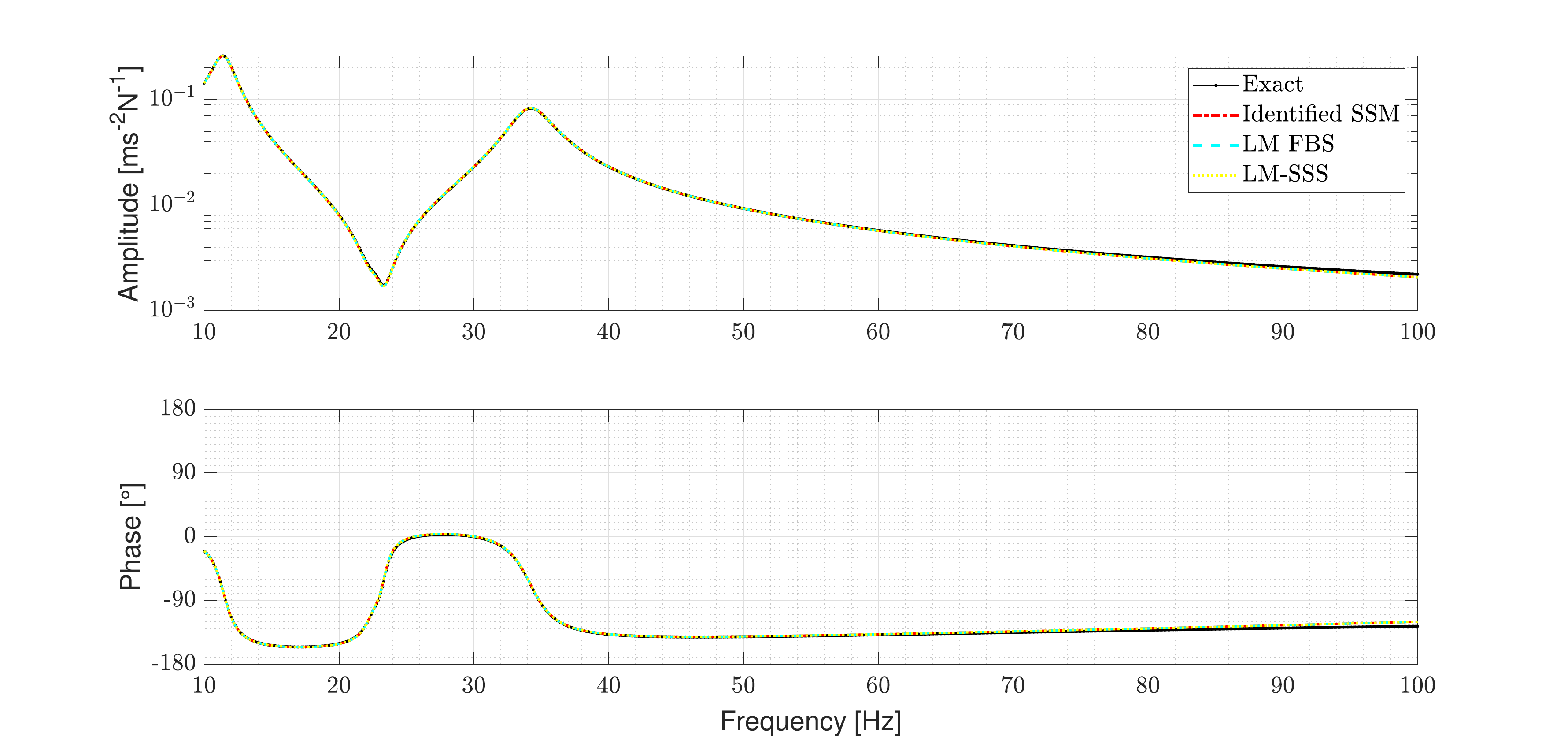}
    \caption{Accelerance FRF of the component $B$, whose output is the DOF $p3$ and the input is the DOF $p4$.}
     \label{fig:Decoupling_system_A from_AB}
\end{figure}

Figure \ref{fig:Decoupling_system_A from_AB} shows the comparison of the identified FRF of component $B$ by using LM FBS-based decoupling and the same FRF of the exact model as well as the FRF of the model identified in section \ref{Identified State-Space Models} and the FRF of the model identified by performing LM-SSS-based decoupling. It is well evident that the FRFs of the identified model of component B are well matching the FRF of the exact model.  Furthermore, the identified FRF obtained by using LM FBS-based decoupling and the one of the identified model of component $B$ obtained in section \ref{Identified State-Space Models} are perfectly fitting the one of the identified model obtained by applying LM-SSS-based decoupling. Hence, we may claim that LM-SSS technique is validated to perform decoupling. 

At this point, we are interested in demonstrating that LM-SSS is able to decouple state-space models previously transformed into coupling form and that by using the procedures described in section \ref{Minimal-order coupled state-space models} we may eliminate from the identified model the redundant states, whose presence is due to the decoupling operation. To perform this demonstration, the identification of a state-space model for component $B$ was performed by decoupling the identified model of $A$ previously transformed into different coupling forms from the minimal order coupled state-space models (which are already obtained in coupling form) computed in section \ref{Coupling Results}. 

\begin{figure}[ht]
\centering
    \includegraphics[width=1\textwidth]{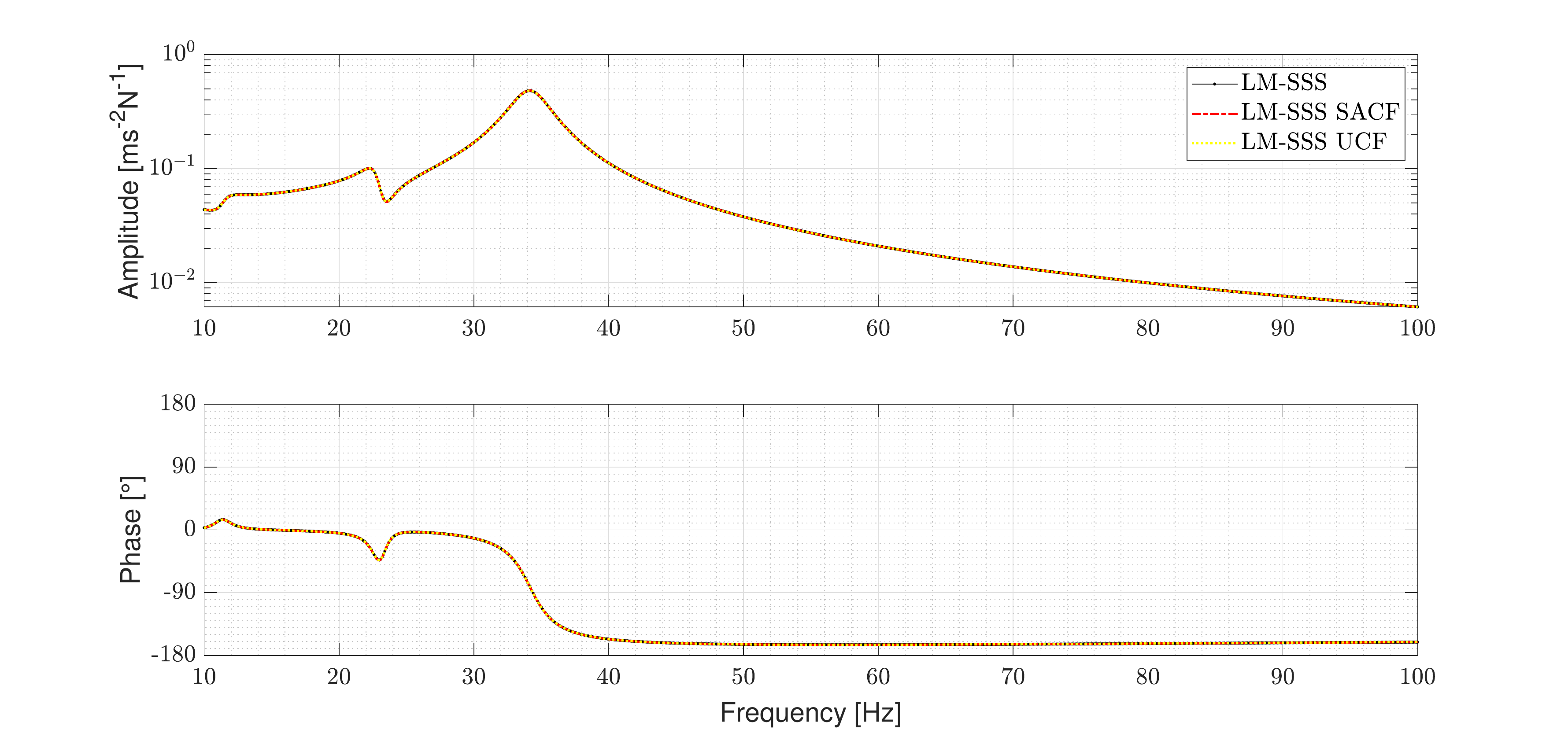}
    \caption{Accelerance FRF of the component $B$, whose output is the DOF $p1$ and the input is the \textcolor{black}{internal} DOF $p3$.}
     \label{fig:Decoupling_system_A from_AB_coupling_form}
\end{figure}

In figure \ref{fig:Decoupling_system_A from_AB_coupling_form} a comparison of a FRF of the identified state-space model of component $B$ by using untransformed models and by using models transformed into different coupling forms is presented. By observing this figure, we may conclude that the FRFs of both identified models are very well matching, validating LM-SSS to decouple state-space models previously transformed into coupling form and the procedures described in section \ref{Minimal-order coupled state-space models} to eliminate the redundant states, whose origin is a consequence of the decoupling operation.

As final analysis, it is worth discussing on the number of states of the identified models of component $B$ by using decoupling. The state-space model obtained by using untransformed models presented $n_{A}+n_{B}+n_{A}=58$ states, while the one obtained by using models transformed into coupling form presented $n_{A}+n_{B}-2n_{J}+n_{A}-2n_{J}=50$ states, being found that all the state-space models presented several unstable poles. It is evident that these models present more states than the identified state-space model obtained from the system identification routines. Hence, the identified models by using LM-SSS decoupling are non minimal-order models. This may explain, together with the lack of passivity of the used state-space models, the presence of unstable poles.

The verified increment of states is a consequence of the inclusion of the dynamics of the component $A$ into the coupled model in order to perform its decoupling (see section \ref{decoupling}). Hence, the dynamics of this component will be present twice on the identified model, since it was already included in the coupled model. Due to the double presence of the dynamics of component $A$, pairs of spurious modes will be present on the identified model \citep{MS_2015}. Distinguishing spurious modes from physical ones might be hard to accomplish, especially, if there is no previous knowledge about the structure to be identified, which is a common situation. Thus, if possible, it is recommended to avoid decoupling operation by using system identification methods to identify state-space models from the FRFs of the components. 

\color{black}

\section{Experimental validation}\label{Experimental validation}

In this section, an experimental validation of the approaches discussed in this paper is provided. We start by presenting in section \ref{Testing Campaign} the mechanical components that were experimentally tested and how these tests were performed. In section \ref{Identified state-space models Experimental} the sets of FRFs obtained from the performed testing campaign are used to estimate state-space models representative of the structures experimentally characterized. To further validate UCF presented in section \ref{Unconstrained_Coupling_Form}, each of the identified state-space models are transformed into the coupling form. Then, the FRFs of the untransformed and of the transformed identified models of each component are compared with the respective measured FRFs. Finally, in section \ref{Decoupling_and_Coupling_Results_Experimental} decoupling and coupling operations are performed with the identified state-space models by using the LM-SSS method (see section \ref{LM_SSS}). To experimentally validate LM-SSS to compute minimal-order coupled models, this technique is also applied to decouple and couple the estimated state-space models previously transformed into UCF, being the elimination of the redundant states present on the computed state-space models performed by following the procedures described in section \ref{Minimal-order coupled state-space models}. The FRFs of the computed coupled state-space model obtained by coupling the untransformed models and of the computed minimal-order coupled model are then compared with the FRFs obtained by applying LM-FBS method and with the respective measured FRFs.

\subsection{Testing Campaign}\label{Testing Campaign}

To obtain experimental data to validate the approaches discussed in this paper, a measurement campaign was performed. This testing camping involved the experimental modal characterization of the following mechanical components:

\begin{itemize}
    \item Two aluminum crosses, from now on labelled as cross aluminum A and B;
    \item Two steel crosses, from now on labelled as cross steel A and B;
    \item Assembled system composed by two aluminium crosses connected by a rubber mount (assembly A);
    \item Assembled system composed by two steel crosses connected by a rubber mount (assembly B).
\end{itemize}

\begin{figure}
\centering
\begin{subfigure}{0.5\textwidth}
  \centering
  \includegraphics[width=0.6\linewidth]{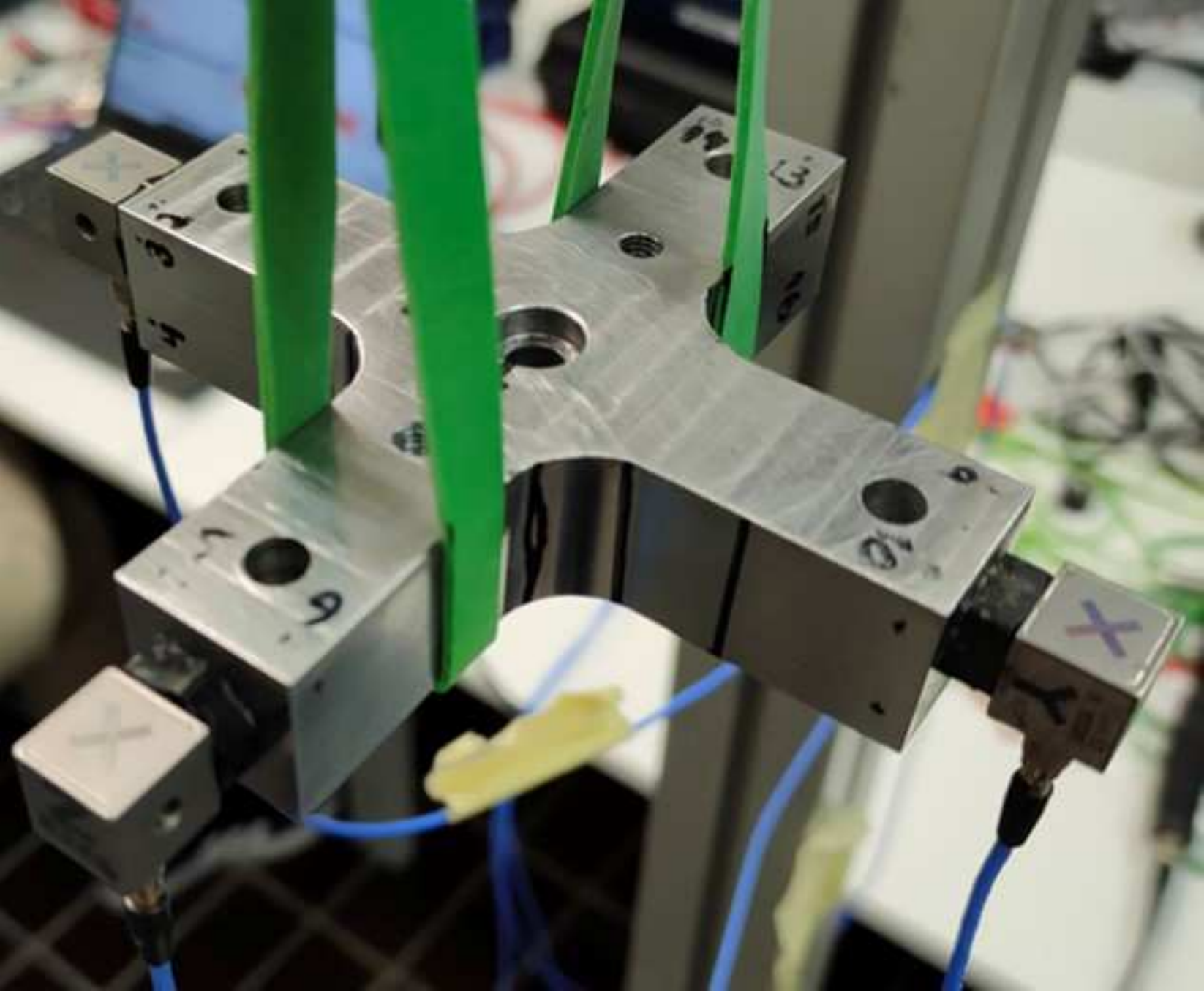}
  \caption{\textcolor{black}{Crosses.}}
  \label{fig:cross}
\end{subfigure}%
\begin{subfigure}{0.5\textwidth}
  \centering
  \includegraphics[width=0.6\linewidth]{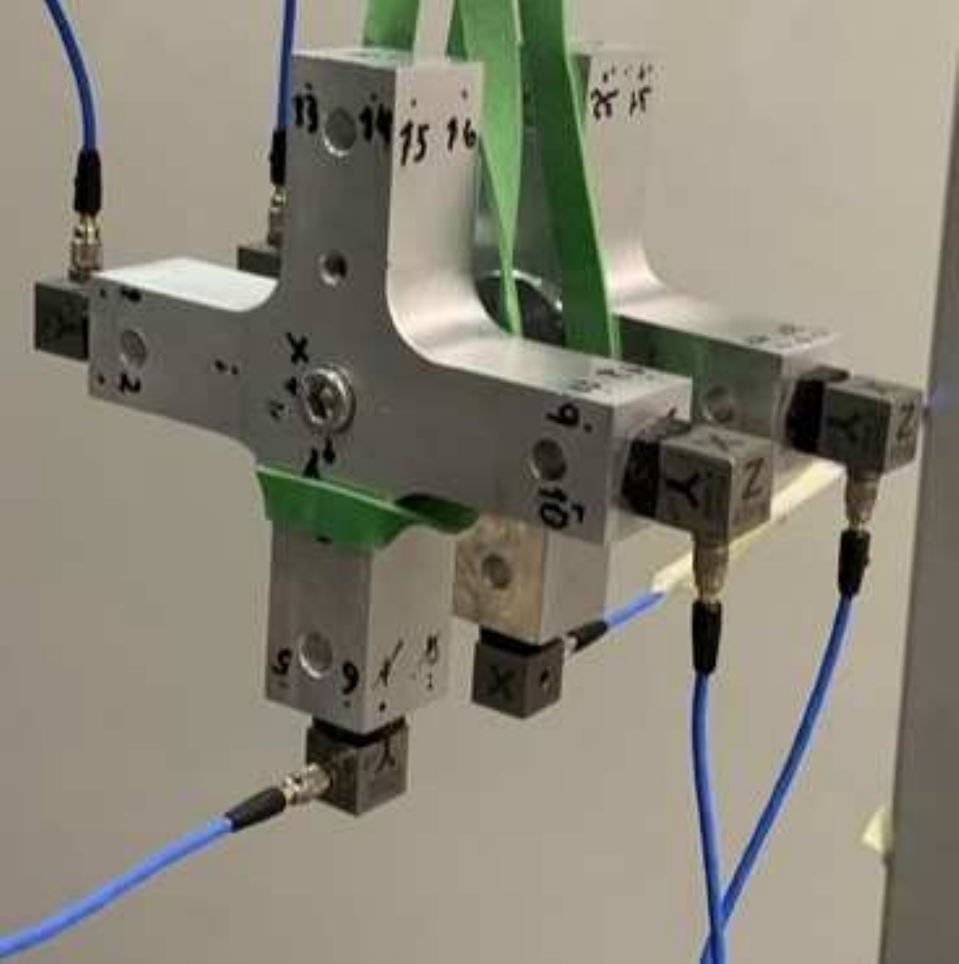}
  \caption{\textcolor{black}{Assemblies.}}
  \label{fig:Assembly}
\end{subfigure}
\caption{\textcolor{black}{Test set up used to experimentally characterize the isolated crosses and assemblies.}}
\label{fig:test}
\end{figure}

\begin{figure}[ht]
\centering
    \includegraphics[width=0.7\textwidth]{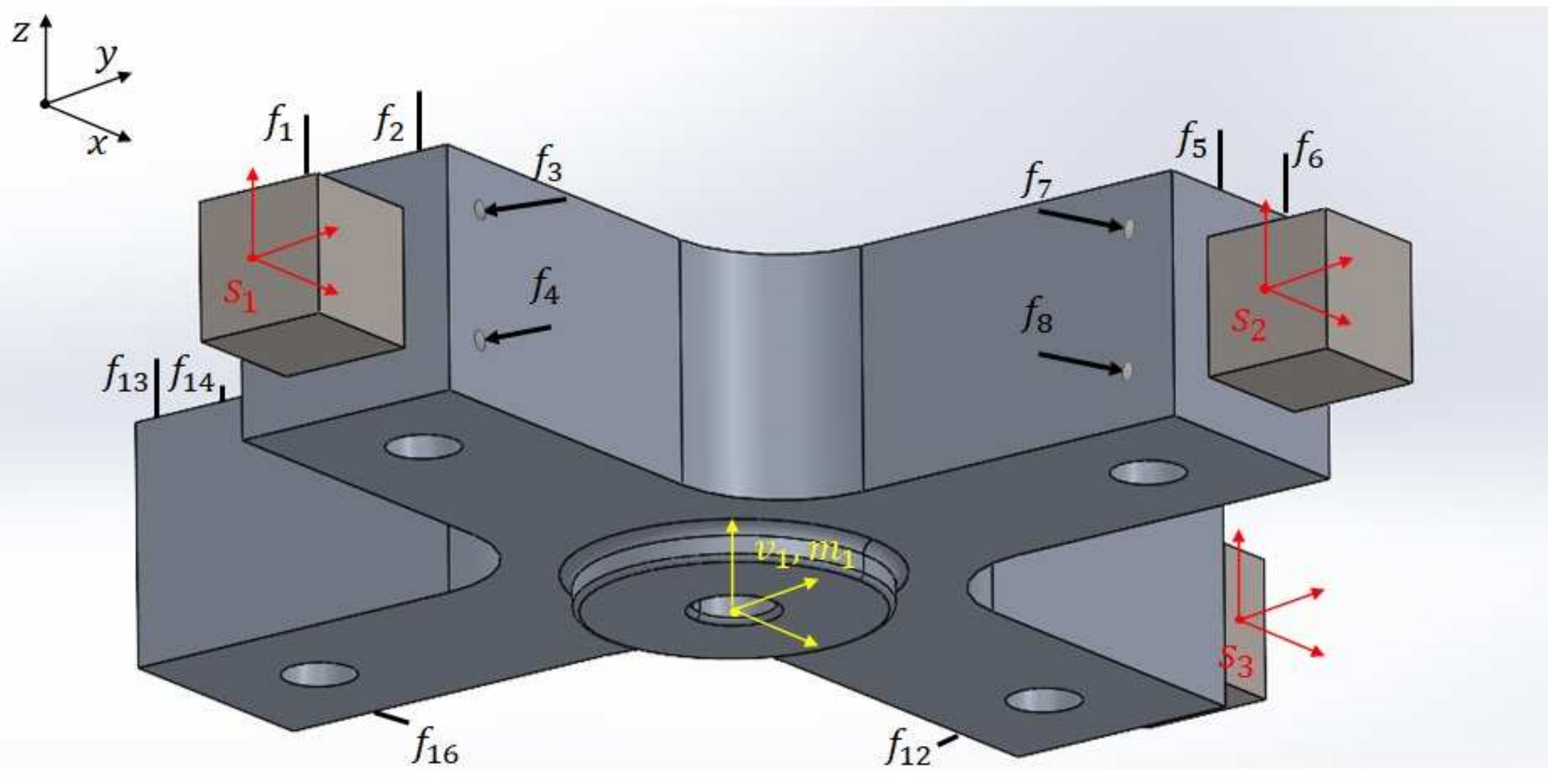}
    \caption{\textcolor{black}{Locations of measurement accelerometers (red), hammer impact directions (black arrows) and virtual point (yellow).}}
     \label{fig:VPT_Cross}
\end{figure}

Figures \ref{fig:cross} and \ref{fig:Assembly} show the test set up used to test the crosses and the assemblies. Rowing hammer mode was chosen as experimental modal analysis testing approach. In fact, each cross, either isolated or included in the assembly, was instrumented with three accelerometers. Hammer impacts were applied at sixteen different locations (see figure \ref{fig:VPT_Cross}). Tests were performed to experimentally characterize each of the four crosses and each of the two assemblies. As measuring and exciting the crosses at interface locations was infeasible, they were manufactured to behave as rigid bodies in the frequency range of interest. In this way, virtual point transformation (VPT) \citep{MV_2013} could be employed to estimate the accelerations and loads at the interface locations (see figure \ref{fig:VPT_Cross}). Furthermore, the use of crosses makes a more effective excitation of the rotational DOFs possible, enabling us to obtain a reliable six DOF characterization of these substructures and, hence a twelve DOF characterization of assemblies A and B. A similar testing campaign and procedure to compute the interface accelerations and loads was reported in \citep{MH_18} and \citep{MH_2020}.

Indeed, the eigenfrequency of the first flexible mode is 5187 $Hz$ for the aluminum cross and 5950 $Hz$ for the steel cross. These eigenfrequencies were obtained by performing a Finite Element analysis of the crosses and by considering the three accelerometers attached to both crosses. The accelerometers were all PCB Model TLD356A32, which are the same sensors as used in \citep{MH_2020}. In that publication, the authors mention that the mass of each accelerometer with cable-connector was measured to be 7.51 g, thus the mentioned eigenfrequencies were calculated by considering that each accelerometer weighted 7.51 g. As the frequency range of interest was defined to be between 20 $Hz$ and 500 $Hz$, the rigid body assumption could be considered valid in the frequency range of interest. 


\subsection{Identified state-space models}\label{Identified state-space models Experimental}

\begin{figure}
  \begin{subfigure}[t]{.45\textwidth}
    \centering
    \includegraphics[width=1\textwidth]{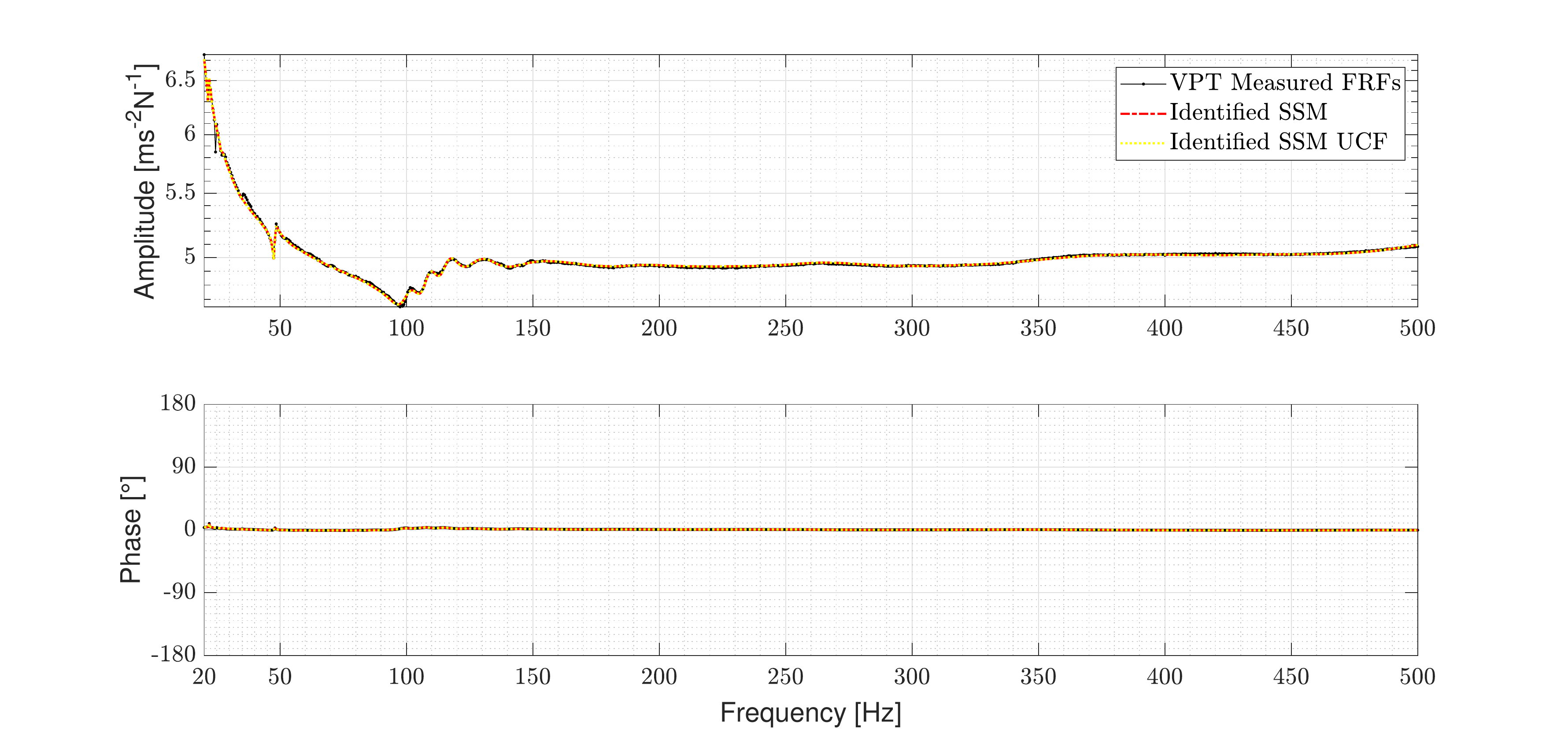}
   \caption{\textcolor{black}{FRF of the aluminum cross B, whose output is the DOF $v_{2}^{z}$ and the input is the DOF $m_{2}^{z}$.}}
     \label{fig:Identified_VPT_Cross_Aluminum}
  \end{subfigure}
  \hfill
  \begin{subfigure}[t]{.45\textwidth}
    \centering
    \includegraphics[width=1\textwidth]{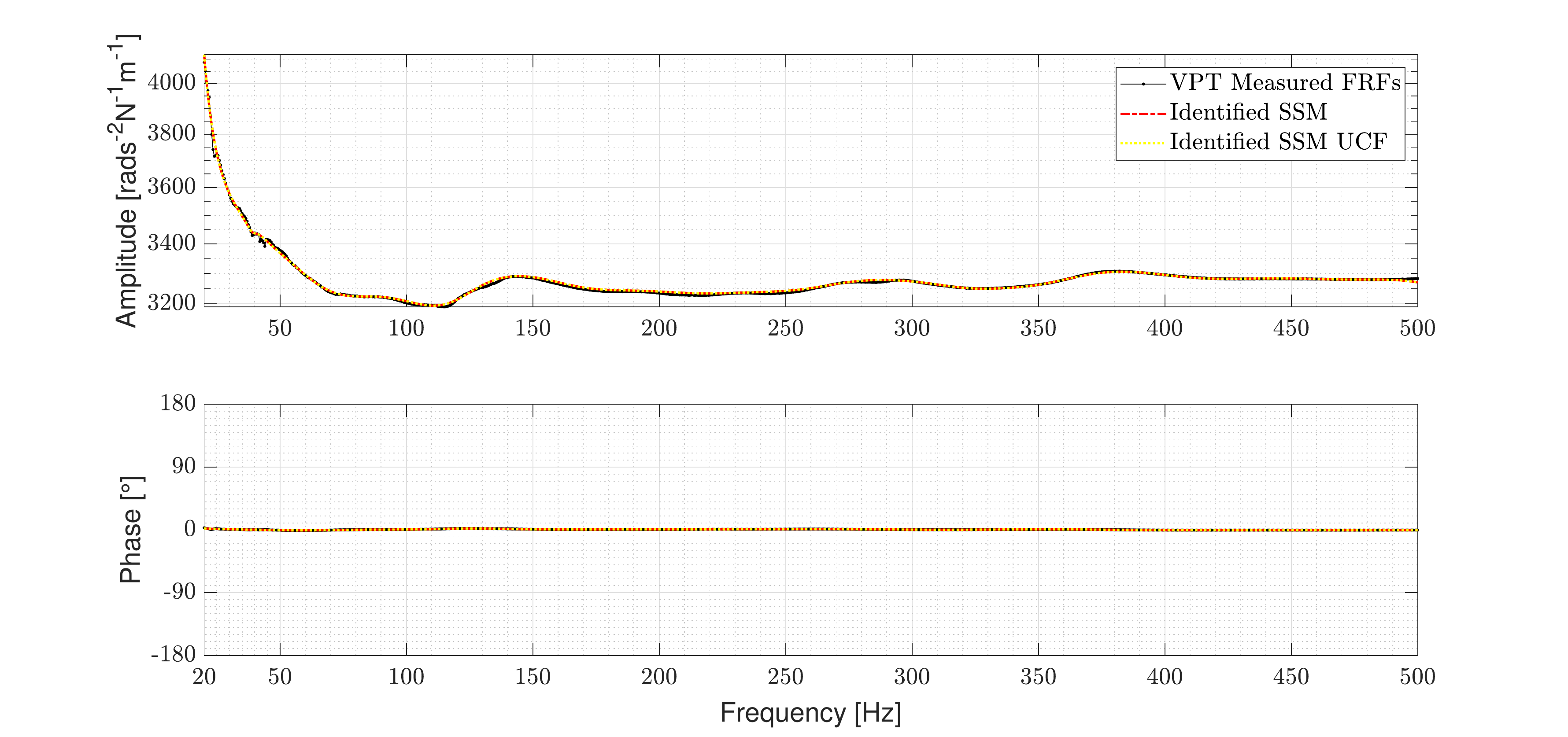}
    \caption{\textcolor{black}{FRF of the steel cross A, whose output is the DOF $v_{1}^{R_{x}}$ and the input is the DOF $m_{1}^{R_{x}}$.}}
     \label{fig:Identified_VPT_Cross_Steel}
  \end{subfigure}

  \medskip

  \begin{subfigure}[t]{.45\textwidth}
    \centering
    \includegraphics[width=1\textwidth]{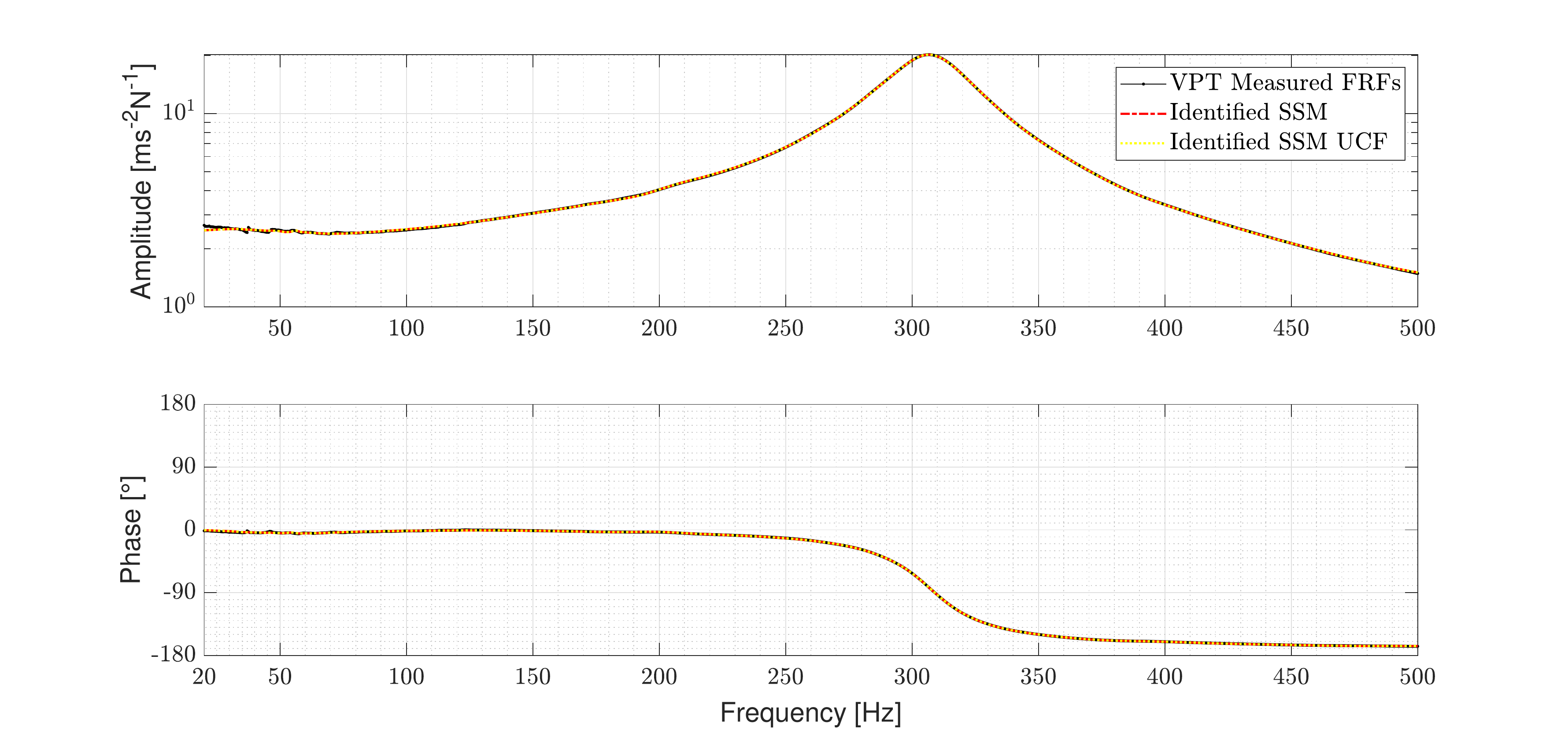}
    \caption{\textcolor{black}{FRF of the assembly A, whose output is the DOF $v_{1}^{z}$ and the input is the DOF $m_{2}^{z}$.}}
     \label{fig:Identified_VPT_Assembly_Aluminum}
  \end{subfigure}
  \hfill
  \begin{subfigure}[t]{.45\textwidth}
    \centering
    \includegraphics[width=1\textwidth]{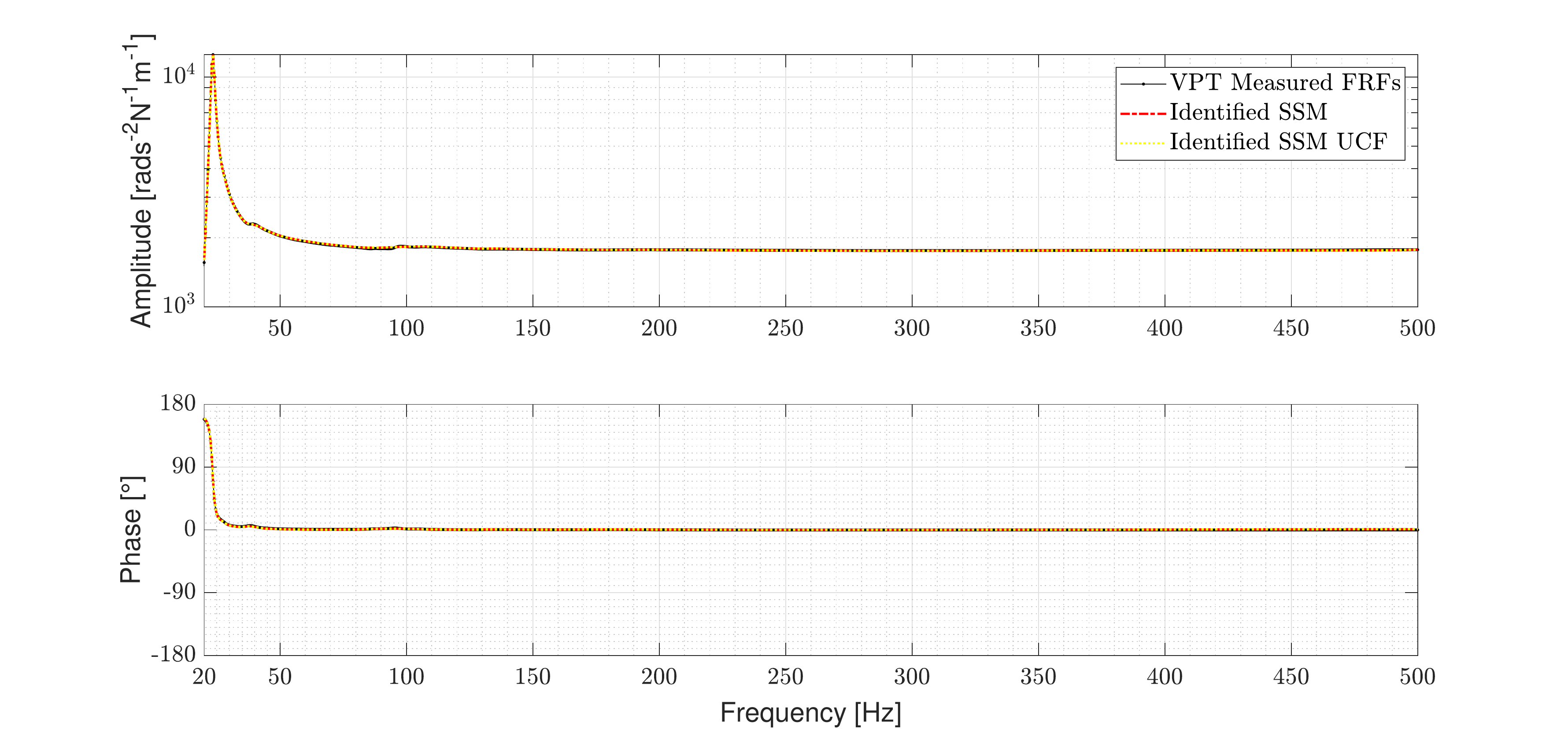}
    \caption{\textcolor{black}{FRF of the assembly B, whose output is the DOF $v_{1}^{R_{z}}$ and the input is the DOF $m_{1}^{R_{z}}$.}}
     \label{fig:Identified_VPT_Assembly_Steel}
  \end{subfigure}
\caption{\textcolor{black}{Comparison of a interface FRF obtained by applying VPT on the measured FRFs with the same FRF of the estimated untransformed state-space models and of the same models transformed into $UCF$ for each of the experimentally characterized structures.}}
  \label{fig:Identified_VPT_State_Space_Models}
\end{figure}

VPT was applied to transform both outputs and inputs of the measured FRFs to the defined virtual points on each cross (placed at the interface of the crosses, see figure \ref{fig:VPT_Cross}). Then, from the computed interface FRFs state-space models representative of each of the experimentally tested components were estimated. Figures \ref{fig:Identified_VPT_State_Space_Models} show, for each mechanical system, the comparison between the FRF obtained by applying VPT on the measured FRFs with the same FRF of the respective estimated state-space model and of the same model transformed into $UCF$. Note that in the caption of figures \ref{fig:Identified_VPT_State_Space_Models} and from now on, the virtual point outputs of the crosses aluminum and steel A will be labelled as $v_{1}$, while for crosses aluminum and steel B these outputs will be tagged as $v_{2}$. The virtual point inputs of crosses aluminum and steel A will be labelled as $m_{1}$, whereas for crosses aluminum and steel B these inputs will be tagged as $m_{2}$.

By observing figure \ref{fig:Identified_VPT_State_Space_Models}, it is clear that the FRFs obtained by applying VPT on the measured FRFs and the FRFs of the estimated state-space models and of the same models transformed into $UCF$ are very well matching for all the structures. Therefore, we may conclude that the system identification method was able to accurately estimate state-space models from experimentally acquired data and that $UCF$ is accurate, even when applied to transform into coupling form state-space models estimated from measured FRFs. 



\subsection{Decoupling and Coupling Results}\label{Decoupling_and_Coupling_Results_Experimental}

\begin{figure}
  \begin{subfigure}[t]{.45\textwidth}
    \centering
    \includegraphics[width=1\textwidth]{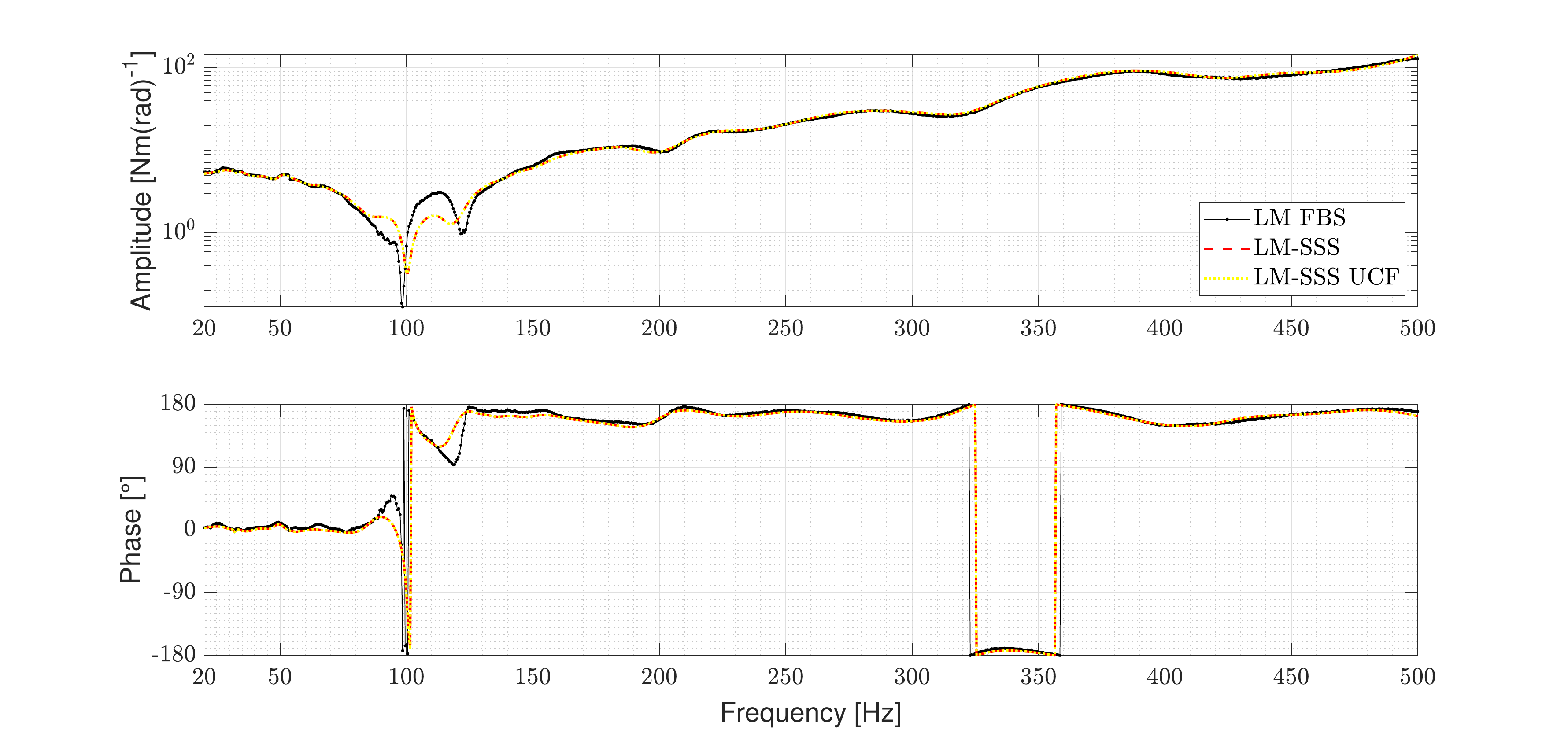}
    \caption{\textcolor{black}{Dynamic stiffness of the rubber mount, whose output is the DOF $v_{1}^{R_{Z}}$ and input is the DOF $m_{1}^{R_{Z}}$.}}
     \label{fig:Z_rubber_mount_rotation}
  \end{subfigure}
  \hfill
  \begin{subfigure}[t]{.45\textwidth}
    \centering
    \includegraphics[width=1\textwidth]{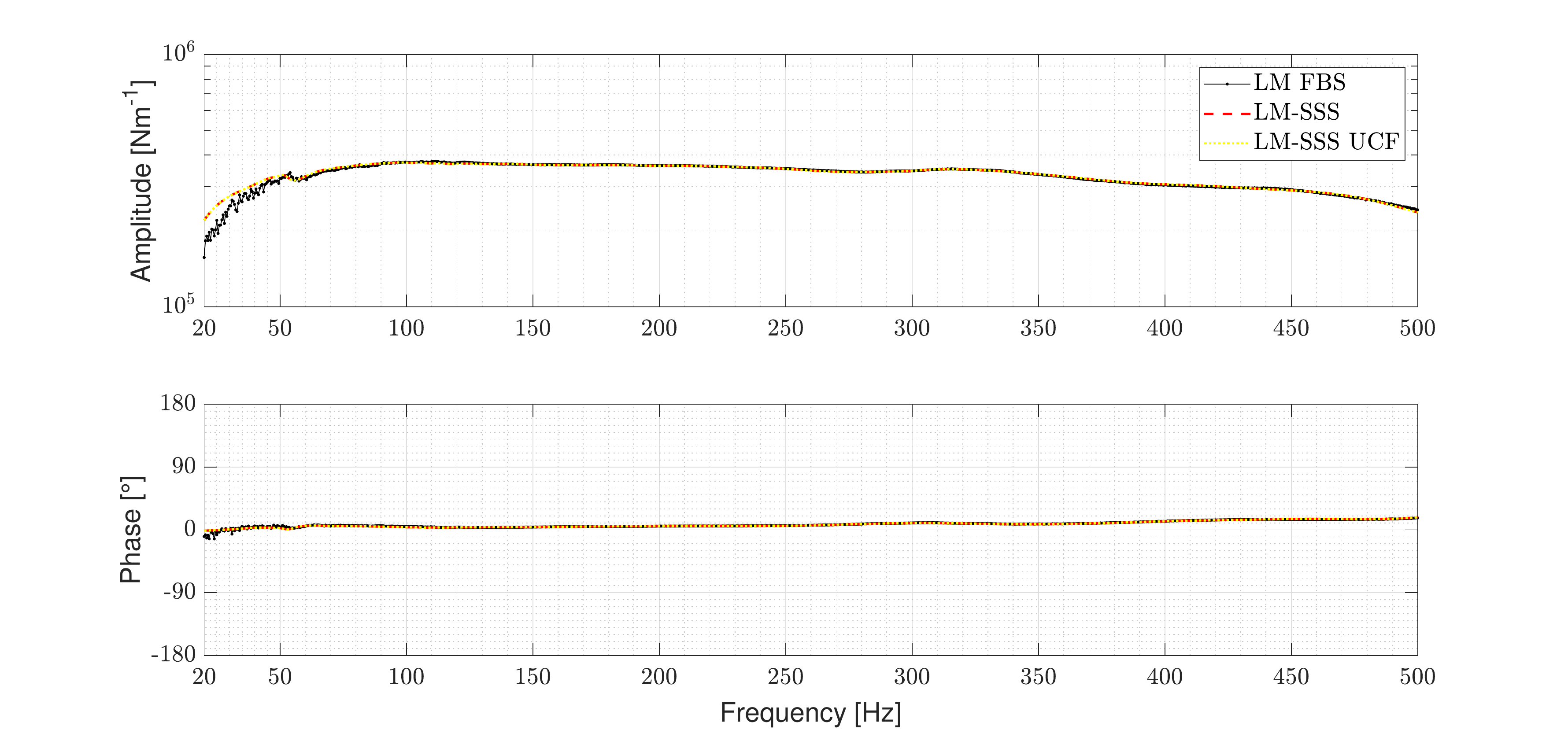}
    \caption{\textcolor{black}{Dynamic stiffness of the rubber mount, whose output is the DOF $v_{2}^{z}$ and input is the DOF $m_{2}^{z}$.}}
     \label{fig:Z_rubber_mount_axial}
  \end{subfigure}
  \caption{\textcolor{black}{Comparison of the dynamic stiffness obtained from the decoupling operation with LM-FBS with the dynamic stiffness of the state-space models obtained by applying LM-SSS to decouple the untransformed state-space models and with the same dynamic stiffness of the minimal-order model obtained by applying LM-SSS to decouple the state-space models transformed into $UCF$ together with the post-processing procedures described in section \ref{Minimal-order coupled state-space models}.}}
  \label{fig:FRFs_identified_rubber_mount}
  \end{figure}

In this section, we will start by applying LM-SSS method to perform the decoupling of both aluminum crosses from the assembly A in order to obtain a state-space model representative of the rubber mount. This operation will be performed with untransformed state-space models and with the same models previously transformed into $UCF$. As the FRFs of the rubber mount are not available, the FRFs obtained from the decoupling operation with the well-known LM FBS technique will be taken as reference. To ease the interpretation of results, the accelerance FRFs obtained from the mentioned approaches are inverted and multiplied by $-\omega^{2}$ to obtain the correspondent dynamic stiffness. Figure \ref{fig:FRFs_identified_rubber_mount} shows a driving point translation and rotation dynamic stiffness obtained by using the following methodologies: i) decoupling operation by applying LM FBS to decouple the interface FRFs of the aluminum crosses obtained by applying VPT on the respective experimentally measured FRFs from the interface FRFs of the assembly A obtained by applying VPT on the respective measured set of FRFs, ii) dynamic stiffness of the state-space model obtained by applying LM-SSS to decouple the untransformed state-space models representative of the aluminum crosses from the one representative of assembly A and iii) dynamic stiffness of the minimal-order model obtained by performing the same decoupling operation as in ii), but with the state-space models previously transformed into $UCF$ and by eliminating the redundant states originated from the decoupling operation by following the procedures described in section \ref{Minimal-order coupled state-space models}.

By observing figure \ref{fig:FRFs_identified_rubber_mount}, it is evident that the dynamic stiffness of the rubber mount identified by LM FBS is very well matching the dynamic stiffness of the state-space model obtained by applying LM-SSS approach to decouple the untransformed models and with the dynamic stiffness of the minimal-order model. Furthermore, the shape of the identified dynamic stiffness is, approximately, a straight line for low frequencies as expected (see, \cite{MH_2020}), which further validates the identified rubber mount dynamic stiffness.  

\begin{figure}
  \begin{subfigure}[t]{.45\textwidth}
    \centering
    \includegraphics[width=1\textwidth]{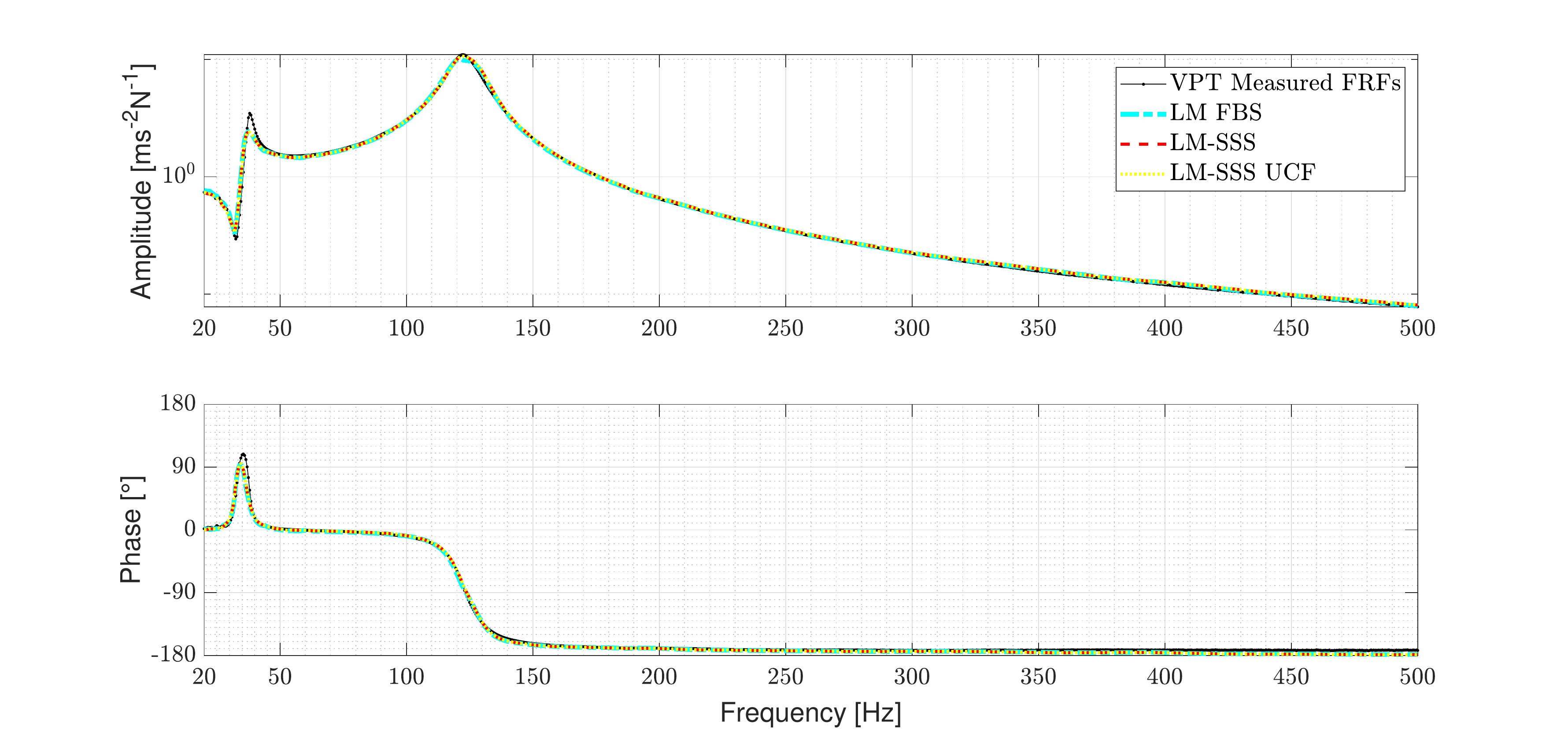}
  \caption{\textcolor{black}{FRF of the assembly B, whose output is the DOF $v_{1}^{x}$ and the input is the DOF $m_{2}^{x}$.}}
  \label{fig:Coupling_Driving_Point_Translation}
  \end{subfigure}
  \hfill
  \begin{subfigure}[t]{.45\textwidth}
    \centering
    \includegraphics[width=1\textwidth]{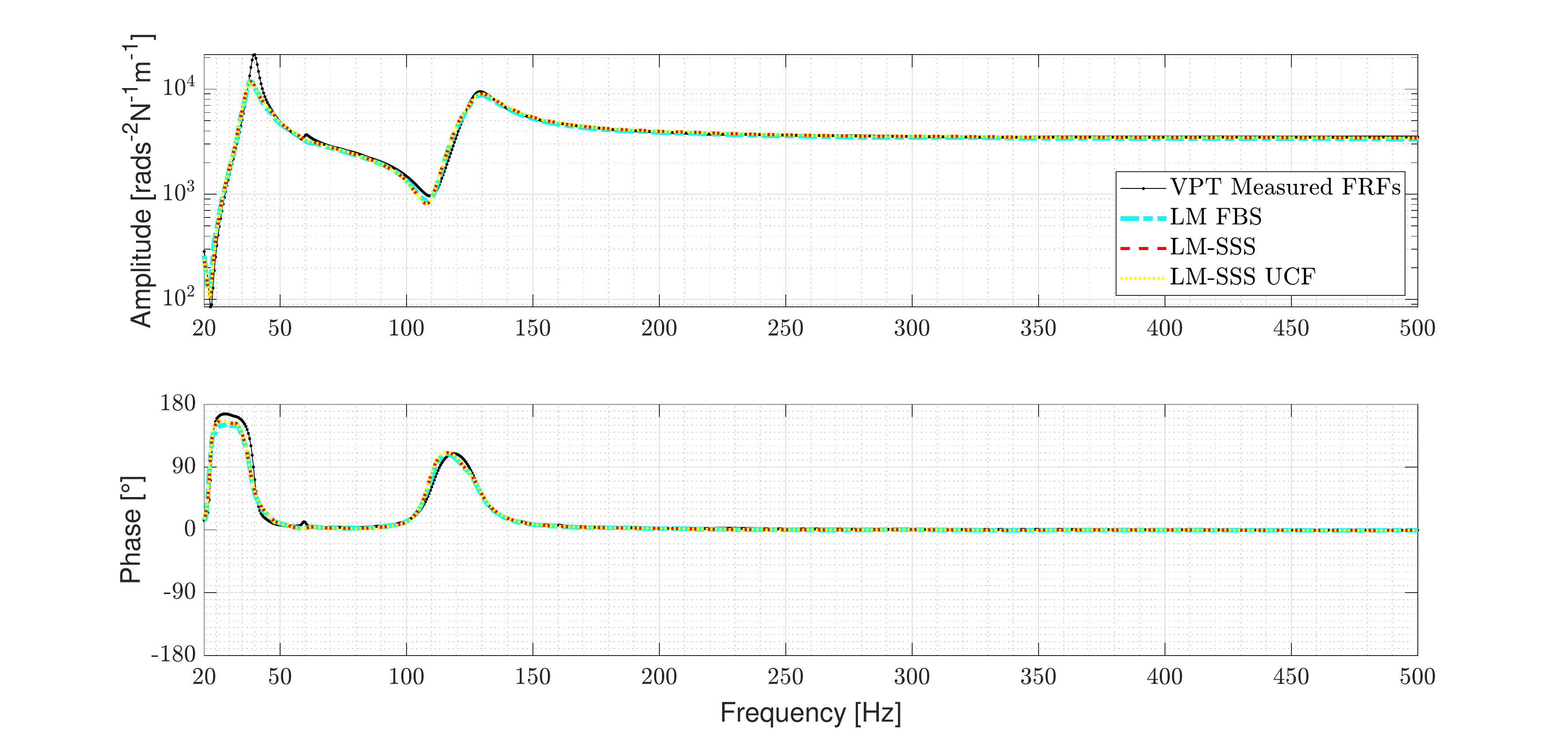}
  \caption{\textcolor{black}{FRF of the assembly B, whose output is the DOF $v_{2}^{R_{x}}$ and the input is the DOF $m_{2}^{R_{x}}$.}}
  \label{fig:Coupling_Driving_Point_Rotation}
  \end{subfigure}
  \caption{\textcolor{black}{Comparison of two interface FRFs of the assembly B computed by applying VPT on its measured FRFs with the coupled FRFs obtained from the coupling operation with LM-FBS, with the FRFs of the coupled state-space model obtained by using LM-SSS to couple the untransformed state-space models and with the FRFs of the minimal-order coupled model obtained by applying LM-SSS to couple the state-space models transformed into $UCF$ together with the post-processing procedures described in section \ref{Minimal-order coupled state-space models}.}}
  \label{fig:Coupling_FRFs_assembly_B}
  \end{figure}

As final analysis, by using LM-SSS the identified state-space models representative of the rubber mount are coupled with the estimated state-space models representative of the steel crosses. Figure \ref{fig:Coupling_FRFs_assembly_B} shows the comparison of two interface FRFs of the assembly B computed by applying VPT on its measured FRFs with the following solutions: i) the same FRFs obtained by coupling with LM FBS the FRFs representative of the rubber mount with the interface FRFs of the steel crosses obtained by applying VPT on the respective measured FRFs, ii) the same FRFs of the coupled state-space model obtained by using LM-SSS to couple the identified rubber mount state-space model and the estimated state-space model representative of the steel crosses and iii) FRFs of the minimal order coupled state-space model obtained by using LM-SSS to couple the identified minimal-order model representative of the rubber mount (which is already obtained in coupling form) with the estimated state-space models of the steel crosses previously transformed into $UCF$ and by following the procedures presented in section \ref{Minimal-order coupled state-space models} to perform the elimination of the redundant states originated from the coupling operation.

By observing figure \ref{fig:Coupling_FRFs_assembly_B}, it is evident that all the solutions are very well matching. Hence, we may claim that LM-SSS method is a reliable method to perform coupling operations with state-space models estimated from measured data. Furthermore, we may conclude that the use of LM-SSS to couple state-space models previously transformed into $UCF$ and the use of the post-processing procedures described in section \ref{Minimal-order coupled state-space models} are valid to compute minimal-order coupled models, even when dealing with state-space models estimated from measured data. 

The coupling operation reported in this section was also performed by using the method proposed by Su and Juang (here labelled as classical SSS). It was found that the FRFs of the computed coupled state-space model were completely equal to the ones of the coupled state-space model obtained by LM-SSS. To better assess the advantage of using LM-SSS instead of classical SSS in terms of computation time, we have measured the time required to compute the matrix inversions required by the classical SSS method and to calculate the single matrix inversion demanded by the LM-SSS method. 
The benchmark test was carried out $10000$ times to give statistical consistency to the calculation time results. The matrix inversions process of classical SSS can take place in $2.12 \times 10^{-5} s$ (average value over the statistical sample the $10000$ trials), while the single matrix inversion of LM-SSS in $5.49 \times 10^{-6} s$ (average value over the statistical sample the $10000$ trials), hence four times faster. The reported analysis was performed on an Intel(R) Core(TM) i7-8550U CPU and 8 GB RAM machine. For the coupling operation discussed in this paper this improvement is not substantial. However, for some applications, e.g. when dealing with real-time substructuring applications involving systems presenting time varying performances, this improvement might result in an important decrease on the computational cost of the application, because several coupling operations are required to be done. Furthermore, for coupling operations with substructures presenting more interface DOFs, it is expected that the benefit in terms of computational time of using LM-SSS instead of classical SSS be even greater, as highlighted in section \ref{Comparison of LM-SSS with other SSS techniques}.

\color{black}

\section{Conclusion}\label{Conclusion}

The LM-SSS method showed to be reliable to couple and decouple state-space models \textcolor{black}{estimated from both numerical and experimental FRFs. Moreover, it has also been demonstrated that LM-SSS is able to couple and decouple state-space models transformed into coupling form (either into $SACF$ or $UCF$) and that the post-processing procedures described in section \ref{Minimal-order coupled state-space models} are accurate to perform the elimination of the redundant states originated from coupling and decoupling operations (see sections \ref{Numerical example} and \ref{Experimental validation})}. It was also shown that the FRFs of the coupled models calculated by LM-SSS are exactly the same as the coupled FRFs obtained by using LM FBS method \citep{DK06} (see section \ref{Coupling Results}). The same applies for the decoupling operation (see section \ref{Decoupling Results}).

Furthermore, due to the LM-SSS coupling formulation it was possible to establish a new coupling form, denoted unconstrained coupling form (see section \ref{Unconstrained_Coupling_Form}). This coupling form holds the advantage of just requiring the computation of a nullspace and does not rely on the choice of a subspace from the computed nullspace. The performance of UCF showed to be similar to the coupling forms already available to the scientific community (see sections \ref{Identified State-Space Models}, \ref{Coupling Results} and \ref{Decoupling Results}). 

\section*{Credit authorship contribution statement}

\textbf{R.S.O. Dias}: Conceptualization, Investigation, Methodology, Software, Formal analysis, Validation, Data curation, Writing - original draft, Writing - review \& editing. \textbf{M. Martarelli}: Conceptualization, Methodology, Resources, Funding Acquisition, Writing - review \& editing, Supervision, Project administration. \textbf{P. Chiariotti}: Conceptualization, Methodology, Resources, Funding Acquisition, Writing - review \& editing, Supervision, Project administration.

\section*{Declaration of Competing Interest}

The authors declare that they have no known competing financial interests or personal relationships that could have appeared to influence the work reported in this paper.

\section*{Acknowledgements}

The authors gratefully acknowledge Dr. Mahmoud El-Khafafy from Siemens Industry Software NV for supporting the research by providing the system identification algorithm to estimate state-space models from FRFs.

\section*{Funding}

This project has received funding from the European Union's Framework Programme for Research and Innovation Horizon 2020 (2014-2020) under the Marie Sklodowska-Curie Grant Agreement nº 858018.


\appendix
\textcolor{black}{\section{New Coupling Form}\label{Appendix}}


In this section, the coupling form presented in \textcolor{black}{\cite[section~4.9.4]{AMRDvMTPATMR_2020}} will be deduced and analyzed in order to demonstrate that there is no guarantee that its transformation matrix will be full rank. Hence, there exists the possibility that the computed transformation matrix be singular, which would lead to ill-conditioned state vector transformations.

A general transformation matrix to transform a state vector into coupling form is given as follows \citep{SJO_20072697}:
 
\begin{equation}\label{eq:TSjZ}
\textcolor{black}{[T_{NCF}]}=\left[\begin{matrix}
C^{J}_{disp}A\\
C^{J}_{disp}\\
N
\end{matrix}
\right]
\end{equation}

where, $N \in \mathbb{R}^{n-2n_{J}\times n}$.

Let us introduce nullspaces $N_{B} \in \mathbb{R}^{n-n_{J}\times n}$ and $N_{C} \in \mathbb{R}^{n-2n_{J}\times n}$ that are given as follows:

    \begin{subequations}\label{zero_zero}
        \noindent
        \begin{tabularx}{\linewidth}{XX}
        \begin{equation}
        [N_{B}][B^{J}]= \left\{\begin{matrix}
            0
            \end{matrix}
            \right\} \label{NB}
        \end{equation}
        &
        \begin{equation}
        \left[\begin{matrix}
        C_{disp}^{J}A\\
        C_{disp}^{J}
        \end{matrix}
        \right][N_{C}]^T= \left\{\begin{matrix}
        0
        \end{matrix}
        \right\} \label{NC}
        \end{equation}
    \end{tabularx}
    \end{subequations}

where, superscripts $J$ denote quantities associated to interface DOFs.

To avoid numerical problems during the performance of such state transformation, \textcolor{black}{$[T_{NCF}]$} must be invertible and, hence full rank. To make sure that this matrix will be full rank independently of the state-space model under study (provided that its input and output matrices are full column rank and full row rank, respectively), the following relation must be verified:

\begin{equation}\label{eq:full_rank_condition}
\left[\begin{matrix}
C^{J}_{disp}A\\
C^{J}_{disp}
\end{matrix}
\right]\left[\begin{matrix}
N
\end{matrix}
\right]^{T}=\left[\begin{matrix}
0
\end{matrix}
\right]
\end{equation}

At this point, we immediately understand that $[N]=[N_{C}]$ would be a perfect choice to make sure that the relation established in equation \eqref{eq:full_rank_condition} is fulfilled (such choice would lead to UCF presented in section \ref{Unconstrained_Coupling_Form}). However, as previously mentioned, the coupling technique presented in \citep{SJO_20072697} requires that $[N]$ is also a nullspace of $[B^{J}]$, which is a condition not necessarily verified by $[N_{C}]$. Thus, the problem can be posed as follows:

\begin{equation}\label{eq:N_computation}
N=N_{A}N_{B}=N_{C}
\end{equation}

where, $[N_{A}]$ is a matrix that is intended to be calculated so that $[N_{A}][N_{B}]$ is as close as possible to $[N_{C}]$. The calculation of $[N_{A}]$ is therefore a suitable task for least-squares. The computation of this solution can be performed by simply inverting $[N_{B}]$ \citep{HA_2013}, resulting in the following expression:

\begin{equation}\label{eq:full_rank_condition_2}
N_{A}=N_{C}N_{B}^{\dag}
\end{equation}

where, superscript $\dag$ represents the pseudoinverse of a matrix.

Since $[N_{B}]$ is the left nullspace of $[B^{J}]$, $[N_{B}]$ is full row rank and its pseudoinverse can be calculated as follows \citep{ML_2007}:

\begin{equation}\label{eq:NB_pseudoinverse}
N_{B}^{\dag}=N_{B}^{T}(N_{B}N_{B}^{T})^{-1}
\end{equation}

where, $[N_{B}]^{\dag}$ works as right inverse of $[N_{B}]$, hence $[N_{B}][N_{B}]^{\dag}=[I]$.

By using equation \eqref{eq:NB_pseudoinverse} , equation \eqref{eq:full_rank_condition_2} can be rewritten as follows:

\begin{equation}\label{eq:NA}
N_{A}=N_{C}N_{B}^{T}(N_{B}N_{B}^{T})^{-1}
\end{equation}

By using equations \eqref{eq:N_computation} and \eqref{eq:NA}, an expression to calculate $[N]$ can be outlined as given below: 

\begin{equation}\label{eq:variableN}
N=N_{C}N_{B}^{T}(N_{B}N_{B}^{T})^{-1}N_{B}
\end{equation}

Due to the use of least-squares to compute $[N]$, equality \eqref{eq:N_computation} is usually not true, being verified a residual difference between $[N_{C}]$ and $[N]$: 

\begin{equation}\label{eq:full_rank_condition_N}
\mu=N_{C}-N_{C}N_{B}^{T}(N_{B}N_{B}^{T})^{-1}N_{B}=N_{C}-N_{C}N_{B}^{\dag}N_{B}
\end{equation}

where, $[N_{B}]^{\dag}[N_{B}] \neq [I]$, since as previously mentioned $[N_{B}]^{\dag}$ acts as the right inverse of $[N_{B}]$. Hence, there is the possibility of this residual difference to invalidate the equality presented in equation \eqref{eq:full_rank_condition}, which could promote the construction of a \textcolor{black}{$[T_{NCF}]$} matrix that is not full rank and, thus not invertible.

Let us develop expression \eqref{eq:full_rank_condition} in order to better understand how the residual difference given by equation \eqref{eq:full_rank_condition_N} can foster the construction of a singular \textcolor{black}{$[T_{NCF}]$} matrix. By using equation \eqref{eq:variableN} and the transpose property $(AB)^{T}=B^{T}A^{T}$ \citep{HA_2013}, $[N]^{T}$ may be established as follows:

\begin{equation}\label{eq:transpose_N}
N^{T}=N_{B}^{T}((N_{B}N_{B}^{T})^{-1})^{T}N_{B}N_{C}^{T}
\end{equation}

The combination of equation \eqref{eq:transpose_N} and equation \eqref{eq:full_rank_condition} leads to:

\begin{equation}\label{eq:full_rank_condition_3}
\left[\begin{matrix}
C^{J}_{disp}AN_{B}^{T}((N_{B}N_{B}^{T})^{-1})^{T}N_{B}N_{C}^{T}\\
C^{J}_{disp}N_{B}^{T}((N_{B}N_{B}^{T})^{-1})^{T}N_{B}N_{C}^{T}
\end{matrix}
\right]=\left[\begin{matrix}
0
\end{matrix}
\right]
\end{equation}

Assuming that $[N_{B}]$ is calculated to be an orthonormal basis for the nullspace of $[B^{J}]$, which is a common practice, $[N_{B}][N_{B}]^{T}=[I]$ \citep{HA_2013} and the following relation can be established:

\begin{equation}\label{eq:Transp_relation_NB}
((N_{B}N_{B}^{T})^{-1})^{T}=(N_{B}N_{B}^{T})^{-1}
\end{equation}

By using equations \eqref{eq:NB_pseudoinverse} and \eqref{eq:Transp_relation_NB}, equation \eqref{eq:full_rank_condition_3} can be rewritten as follows:

\begin{equation}\label{eq:full_rank_condition_final}
\left[\begin{matrix}
C^{J}_{disp}AN_{B}^{\dag}N_{B}N_{C}^{T}\\
C^{J}_{disp}N_{B}^{\dag}N_{B}N_{C}^{T}
\end{matrix}
\right]=\left[\begin{matrix}
0
\end{matrix}
\right]
\end{equation}

Equality \eqref{eq:full_rank_condition_final} may not be verified, because $[N_{B}]^{\dag}[N_{B}] \neq [I]$. Hence, we may conclude that $[N]$ will not necessarily verify expression \eqref{eq:full_rank_condition}, being impossible to guarantee that \textcolor{black}{$[T_{NCF}]$} will be full rank, independently of the state-space model intended to be transformed.

 \bibliographystyle{elsarticle-num} 
 \bibliography{cas-refs}





\end{document}